\documentclass[%
 reprint,
 amsmath,amssymb,
 aps,
]{revtex4-2}

\usepackage{graphicx}
\usepackage{dcolumn}
\usepackage{bm}
\usepackage{lipsum} 

\begin{document}

\preprint{APS/123-QED}

\title{Constraining Axion-to-Nucleon interaction via ultranarrow linewidth in the Casimir-less regime }%

\author{Lei Chen}
 \altaffiliation[Also at ]{Key laboratory of Artificial Structures and Quantum Control (Ministry of Education),School of Physics and Astronomy, Shanghai Jiao Tong University, 800 Dong Chuan Road, Shanghai 200240, China}
 \author{Jian Liu}
 \altaffiliation[Also at ]{Key laboratory of Artificial Structures and Quantum Control (Ministry of Education),School of Physics and Astronomy, Shanghai Jiao Tong University, 800 Dong Chuan Road, Shanghai 200240, China}
\author{Kadi Zhu}%
 \email{zhukadi@sjtu.edu.cn}
\affiliation{Key laboratory of Artificial Structures and Quantum Control (Ministry of Education),School of Physics and Astronomy, Shanghai Jiao Tong University, 800 Dong Chuan Road, Shanghai 200240, China
}%

\begin{abstract}
 In this paper we develop a quantum optical method to detect the axion-nucleon interaction. We ultilize a levitated optomechanical system consisting of a silica nanosphere and an optical cavity here.  We translate the trapping positions of the nanosphere,  resulting the shift of its resonance frequency, which can be determined from measuring the resulting resonance shift in the transmission spectrum. Furthermore, The frequency shift can be related to the additional forces due to two-axion exchange via substraction. Based on noise ananlysis, estimation and calculation, we set the stringent prospective constraints for the coupling constants of axion-neucleon interaction $g_{an}$ and $g_{ap}$. In the case of $g_{an}^2=g_{ap}^2$, our constraints are most stringent at an ultrawide axion mass range approximately from $10^{-4}\mu eV$ to $10$ $eV$.

\end{abstract}

\maketitle


\section{\label{sec:level1}Introduction}
Axion as a  new light pseudoscalar particle was predicted in 1978 \cite{PhysRevLett.40.223, PhysRevLett.40.279} 
. Since then, it remains the most compelling solution to the strong-CP problem in QCD and a well-motivated dark matter candidate \cite{2014Experimental, PhysRevD.86.010001, PhysRevD.98.030001, Masahiro2013Axions} 
. Due to this, a host of ultrasensitive experiments have been conducted to search for axions and axion-like paricles (ALPs) \cite{2014Experimental, 2018New, 2018Constraining, 2017Search, Leslie2000Searches, 2020Extended}. Since one main property of axion is that it can interact with nucleons \cite{2018New, 2021Invisible}, the axion-to-nucleon interaction has been focused on by amounts of work (see reviews \cite{2017Recent, 2020The}). Consequently,  lots of effective constraints \cite{PhysRevD.101.056013, 2020The} on the coupling constant $g_{an}$ over a wide axion mass range have been eatablished.  Vasilakis et al. set constraints at $ 10^{-4} eV <m_a <1\mu eV$ via magnetometer measurements \cite{2009Limits}. An upper bound  most stringent at $1\mu eV< m_a <1.7meV$ was derived in \cite{2006Particle} by ultilizing the data from a torsion-balance search for Yukawa violations of the gravitational inverse-square law \cite{2007Tests1}.
The strongest constraints at about $1meV < m_a <0.5eV$  were derived in \cite{2015Improved}  from the measurement results of a Casimir-less experiment  \cite{2016Stronger}. 
Several upper bounds \cite{2014Constraining, 2014Constraints, 2017Constraints, V2014Stronger, V2014Constraints, PhysRevD.101.056013} are derived from measuring some Casimir-effect-based objects including effctive Casimir pressure \cite{2007Tests,2010Novel}, the lateral Casimir force between corrugated surfaces \cite{2009Demonstration,2010Lateral}, the difference in Casimir forces \cite{2016Isoelectronic},
the gradient of the Casimir force \cite{2012Gradient} , the Casimir-Polder force \cite{2007Measurement} and  the Casimir force in nanometer separation range \cite{PhysRevB.93.085434}.
The most stringent constraints at $m_a>0.5eV$ were obtained from experiments on measuring the forces between protons in the beam of molecular hydrogen \cite{1979The, 2013Constraints}. The strongest laboratory limits at $m_a>200eV$ were obtained from the experiment on nuclear magnetic resonance \cite{2013Constraints}. Though these effective constraints have been established, it is still desirable for us to search for the axion-to-nucleon interaction and set stronger constraints on it.

In this paper, we ultilize an optomechanical system consisting of a silica nanosphere and an optical cavity which is composed of a mirror and source mass to detect the axion-neucleon interaction. We trap the nanosphere near the surface of the source mass, causing the  resonance frequency of the nanosphere perturbed by the total force gradient including the additonal force due to two-axion exchange. Moreover, we translate the positions of the optical trap, causing the variation of the perturbed resonace frequency. Via subtraction, we eliminate the effct of Casimir interaction and then derive the ralationship between the difference of the perturbed resonace frequencies and the additional forces. With a pump laser and a probe one applied, this difference can be converted to a resonace shift in the transmissin spectrum. By measuring this resonance shift, the value of the difference can be determined. Further, via resonable estimation and numerical calculation, we set proepective constraints on the coupling constants $g_{an}$ and $g_{ap}$. In the case of $g_{an}^2=g_{ap}^2$, our constraints on $g_{an}$ improve on existing bounds by several orders of magnitude at about $10^{-4}\mu eV< m_{a}<10 e$ $V$. Finally, we expect our scheme can be realized in the relevant experiments.

The remainder of the paper is organized as follows: In Sec. II we describe the theoretical model which is based on levitated cavity optomechanics, in Sec. III we present our detction priciple,
in Sec. IV we perform the noise analysis and set the prospective constraints, in Sec. V we summarize the paper.
 
\section{\label{sec:level1}Theoretical model}

\begin{figure}
   \includegraphics[width=25em]{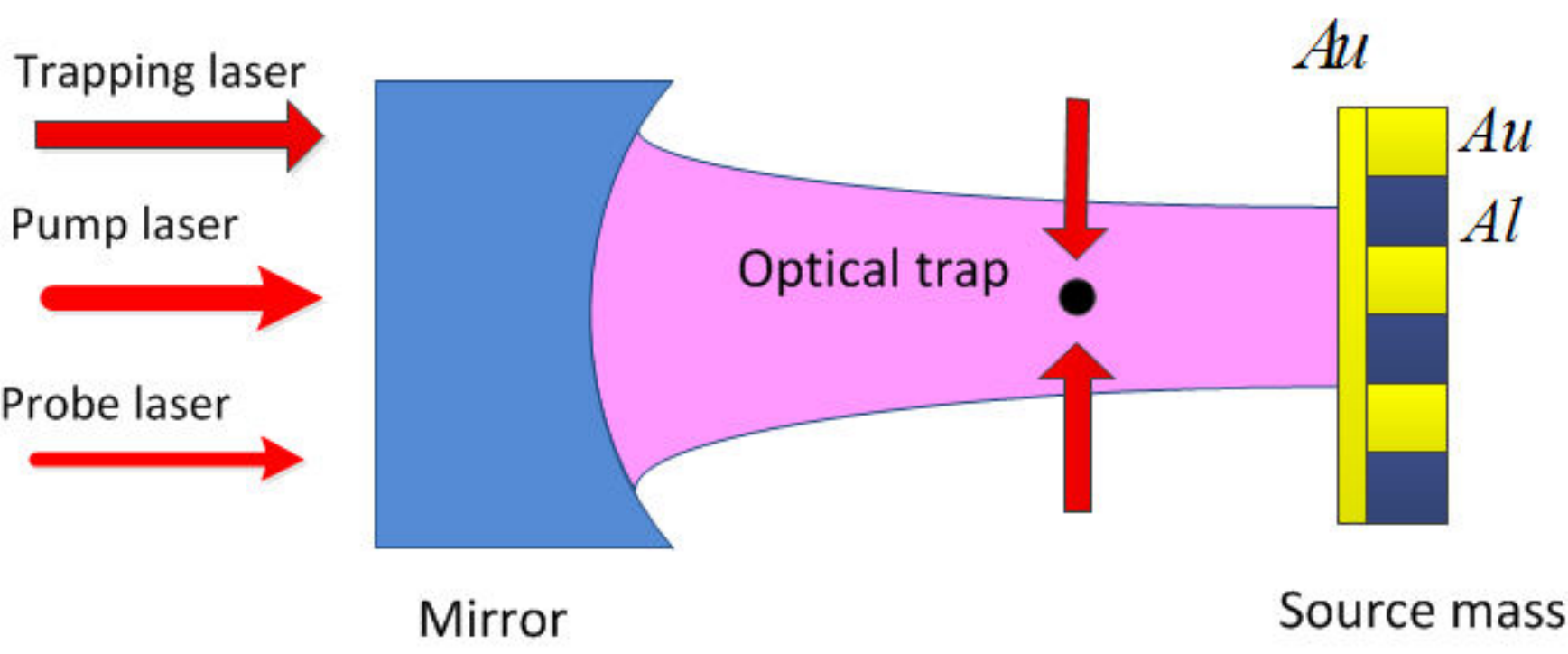}
	\caption{\label{fig:epsart}Schematic setup. A silica nanosphere is trapped in a cavity consisting of a mirror and source mass. The source mass is composed of alternating sections of Au and Al coated with a layer of Au.}
\end{figure}
Here we consider a levitated optomechanical system which is composed of a optical cavity and a silica nanosphere. The cavity consists of a mirror and source mass , which is made of alternating sections of Au and Al coated with a layer of Au (see Fig. 1). The nanosphere is cooled and trapped in the cavity as shown in Fig. 1. Applying a pump laser and a probe laser, the Hamiltonian of this system can be written as \cite{2008Ground,2013Optomechanics, 2020Optomechanics}
\begin{align}
H=&\hbar \omega_m b^+ b +\hbar \omega_c a^+ a+\hbar g (b^+ +b)a^+ a\notag\\
  &+i\hbar E_{pu}( a^+ e^{-i\omega_{pu} t}- a e^{i\omega_{pu} t})\notag \\
  &+i\hbar E_{pr}( a^+ e^{-i\omega_{pr} t}- a e^{i\omega_{pr} t}),
\end{align}
where $\omega_{m}$ is the resonance frequency of the nanosphere resonator and $b^+$ ($b$) is the corresponding  creation (annihilation) operator, 
$\omega_{c}$ is the resonant frequency of the cavity and $a^+$ ($a$) is the corresponding  creation (annihilation) operator, $g$ characterizes the coupling strength between the cavity and the nanosphere, $\omega_{pu}$ and $\omega_{pr}$ are the frequnecies of the pump laser and probe laser respectively,  $E_{pu}$ and $E_{pr}$ are related to the laser power $P$ by $|E_{pu}|=\sqrt{2P_{pu}\kappa/\hbar\omega_{pu}}$ and 
$|E_{pr}|=\sqrt{2P_{pr}\kappa/\hbar\omega_{pr}}$
respectively, where $\kappa$ is the decay rate of the cavity amplitude.

In a rotating frame at a driving field frequency $\omega_{pu}$, the Hamiltonina can be transformed to 
\begin{align}
\tilde{H}=&\hbar \omega_m b^+ b +\hbar \Delta a^+ a+\hbar g (b^+ +b)a^+ a\notag\\
  &+i\hbar E_{pu}( a^+-a)
  +i\hbar E_{pr}( a^+ e^{-i\delta t}- a e^{i\delta t}) ,
\end{align}
where $\delta=\omega_{pr}-\omega_{pu} $ and $\Delta=\omega_{c}-\omega_{pu}$. 
Defining $\tau\equiv \frac{b+b^{+}}{\sqrt{2}}$, and applying Heisenberg equation of motion, we obtain
\begin{equation}
	\frac{da}{dt}=-i\Delta a -ig(b^{+}+b)a+E_{pu}+E_{pr} e^{-i\delta t},
\end{equation}
and 
\begin{equation}
	\frac {d^2 \tau}{dt^2} +\omega_m^2 \tau =-\sqrt{2}g \omega_m a^{+} a.
\end{equation}
Taking the damping terms into consideration, Eqs. (3)-(4) can be rewritten as 
\begin{equation}
	\frac{da}{dt}+(i\Delta+\kappa) a = -ig(b^{+}+b)a+E_{pu}+E_{pr} e^{-i\delta t},
\end{equation}
and 
\begin{equation}
	\frac {d^2 \tau}{dt^2} +\gamma_m \frac {d \tau}{dt}+ \omega_m^2 \tau =-\sqrt{2}g \omega_m a^{+} a,
\end{equation}
where $\gamma_m$ is the damping rate of the mechanical resonator.
Taking  expectation values of Eqs.(5)-(6), we obtain 

\begin{equation}
\langle \frac{da}{dt} \rangle+(i\Delta+\kappa) \langle a\rangle = -ig\sqrt{2}\langle \tau a \rangle+E_{pu}+E_{pr} e^{-i\delta t},
\end{equation}
and
\begin{equation}
	\frac {d^2 \langle \tau \rangle}{dt^2} +\gamma_m \frac {d  \langle \tau \rangle}{dt}+ \omega_m^2  \langle \tau \rangle =-\sqrt{2}g \omega_m \langle a^{+} a\rangle.
\end{equation}
We make the ansatz as  follows:
\begin{equation}
\langle a(t) \rangle= a_0 +a_+ e^{-i\delta t}+ a_- e^{i\delta t},
\end {equation}

\begin{equation}
\langle \tau(t) \rangle= \tau_0 +\tau_+ e^{-i\delta t}+ \tau_- e^{i\delta t}.
\end {equation}
Also, we assume that
\begin{equation}
\langle a^+ a \rangle =\langle a^+\rangle \langle a\rangle ,	
\end{equation}
and
\begin{equation}
	\langle \tau a \rangle = \langle \tau \rangle \langle a\rangle .	
\end{equation}
Substituting Eqs. (9)-(12) into Eq. (7) and (8) respectively and then performing some calculations, we finally attain
\begin{equation}
	|E_{pu}|^2=[\kappa^2+(\Delta-\frac{2g^2 \sigma}{\omega_m})^2]	\sigma,
\end{equation}
where $\sigma$ is defined as $\sigma \equiv |a_0|^2$,
and
\begin{equation}
	a_+ =\frac{E_{pr} K_1 (K_1 K_2-iK_4)}{(K_1 K_3-iK_4)(K_1 K_2-iK_4)+K_4^2},
\end{equation}
with
\begin{align}
K_1 =&\omega_m^2-i\delta\gamma_m-\delta^2,\notag\\
K_2 =&-\kappa+i\delta+i\Delta-\frac{2ig^2\sigma}{\omega_m},\notag\\
K_3 =&\kappa-i\delta+i\Delta-\frac{2ig^2\sigma}{\omega_m},\notag\\
K_4 =&2g^2\sigma\omega_m.
\end{align}

To investigate the optical property of the output field for our system, using an input-output relation, which is valid for a one-sided open cavity: $a_{out}(t) =a_{in}(t)-\sqrt{2\kappa}a(t)$, where $a_{in}$ and $a_{out}$ are the input and output operators, respectively, we can obtain the expectation value of the output field as 
\begin{align}
	\langle a_{out}(t)\rangle =&(E_{pu}/\sqrt{2\kappa}-\sqrt{2\kappa} a_0)e^{-i\omega_{pu}t}\notag\\& +(E_{pu}/\sqrt{2\kappa}-\sqrt{2\kappa} a_+)e^{-i(\omega_{pu}+\delta)t}\notag\\&-\sqrt{2\kappa}a_-e^{-i(\omega_{pu}-\delta)t}.
\end{align}
The transmission of the probe beam, defined as the ratio of the output and input field amplitudes at the probe frequency is given by \cite{2010Optomechanically}

\begin{equation}	
t= \frac{E_{pr}/\sqrt{2\kappa}-\sqrt{2\kappa}a_+}{E_{pr} /\sqrt{2\kappa}}=1-2\kappa a_+/E_{pr}.
\end{equation}	

In summary, in this section we discribe a levitated optomechanical system and derive the expression of the transmission. Next we demonstrate our detection principle.

\section{\label{sec:level1}detection principle}

\begin{figure}
	\includegraphics[width=13em]{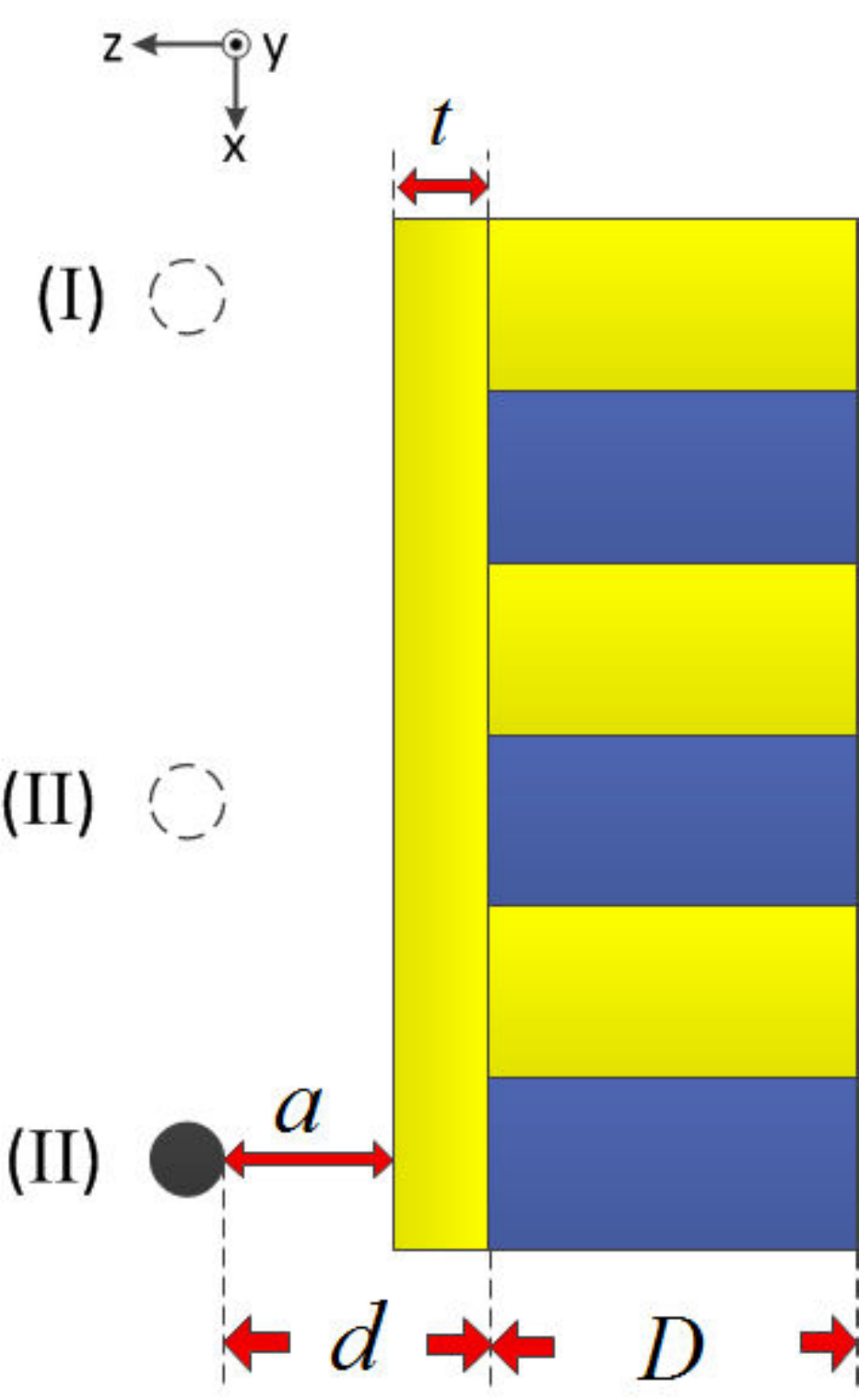}
	\caption{\label{fig:epsart} Schematic diagram near the source mass. The nanosphere is trapped near the source mass with a speration $a\sim 300nm$. The positions of optical trap are classified onto type (I) and type (II). }
\end{figure}
Let us focus on our  proposed system. The radius of the nanosphere is $R\sim 10nm$. The width of the alternatiing sections is $D\sim 100\mu m$ and the thickness of the Au layer is $t\sim200nm$ (see Fig. 2). If we trap the nanosphere near the source mass with a seperation  $a\sim 300nm$, the resonance frequency of the nanosphere will be modified by the total force gradient acting on it. More explicitely, there is
\cite{2012Gradient, 2019Precision, Franz2003Advances}
\begin{equation}
	\frac{\omega^{\prime} - \omega_0}{\omega_0} = -\frac1{2m_s\omega_0^2} \frac{\partial{F_{tot}(d)}}{\partial{d}},
\end{equation}
where $d(\sim 0.5\mu m)=a+t $ is the distance between the rim of the nanosphere and the alternatiing sections,  $\omega_0$ is the unperturbed resonance frequency of the nanosphere while $\omega^{\prime}$ is the modified frequency in the presence of the total force $F_{tot}(d)$, and $m_s$ is the mass of the nanosphere  calculated as $m_s=1.05\times 10^{-20}kg$.

In our scheme, we translate the position of the optical trap along the surface of the source mass, while the separation between the nanosphere and the surface is kept as $a\sim 300nm$.  Note that here among all the alternative sections the specific one which is closest to the nanosphere may be a Au (Al) section. Based on this, we designate the trapping positions where this specific section is Au (Al) as type (I) ((II)), just as shown in Fig. 2. Consequently, all the trapping positions can be classified into these two types. According to Eq. (18), for the trapping positions (I) and (II), the  pertubation of resonance frequency of the nanosphere can be  expressed as

\begin{equation}
	\frac{\omega^{\prime}_{Au} - \omega_0}{\omega_0} = -\frac1{2m_s\omega_0^2} \frac{\partial{F_{tot}^{Au}(d)}}{\partial{d}},
\end{equation}
and
\begin{equation}
	\frac{\omega^{\prime}_{Al} - \omega_0}{\omega_0} = -\frac1{2m_s\omega_0^2} \frac{\partial{F_{tot}^{Al}(d)}}{\partial{d}},
\end{equation}
respectively, where $\omega^{\prime}_{Au}$( $\omega^{\prime}_{Al}$) is the modified resonance frequency, and $F_{tot}^{Au}(d)$  ($F_{tot}^{Al}(d)$) is the total force exerted on the nanosphere.  Note that generally $F_{tot}^{Au}(d)$ and $F_{tot}^{Al}(d)$ are two attractive forces both of which diminish as $d$ increases. Based on this  we conclude that both $\omega^{\prime}_{Au}$ and $\omega^{\prime}_{Al}$ are slightly smaller than $\omega_0$.

Subtracting Eq. (20) from (19), we obtain

\begin{equation}
	\frac{\omega^{\prime}_{Au} - \omega^{\prime}_{Al}}{\omega_0} = \frac1{2m_s\omega_0^2} \frac{\partial\big[F_{tot}^{Al}(d)-F_{tot}^{Au}(d)]}{\partial{d}}.
\end{equation}
We focus on the differential force $F_{tot}^{Al}(d)-F_{tot}^{Au}(d)$. Here we use the Casimir-less technology \cite{2005Constraining, 2016Stronger} to sustract the Casimir background. In detail, we coat the  the alternative sections with a Au layer having a thickness $t\sim200nm$, such that
the difference in the Casimir interaction between the alternative sections with the nanosphere can be significantly attenuated. Consequently, we can suppose the difference of
$F_{tot}^{Al}(d)$ and $F_{tot}^{Au}(d)$ is contributed mainly from the differential  additional force due to two-axion exchange, i.e.,
 
\begin{equation}
	 F_{tot}^{Al}(d)-F_{tot}^{Au}(d)\approx F_{add}^{Al}(d)-F_{add}^{Au}(d),
\end{equation}
where $F_{add}^{Au}(d)$ ($F_{add}^{Al}(d)$) is the additional force acting on the nanosphere trapped at the position (I)((II)). Substituting Eq. (22) into Eq. (21), we obtain

\begin{equation}
	\frac{\omega^{\prime}_{Au} - \omega^{\prime}_{Al}}{\omega_0} \approx \frac1{2m_s\omega_0^2} \big[\frac{\partial F_{add}^{Al}(d)}{\partial{d}}-\frac {\partial F_{add}^{Au}(d)}{\partial{d}}].
\end{equation}

Now we attempt to derive the expression of the difference of  force gradient in Eq. (23). In the system of natural units with $\hbar=c=1$ , the effective potential due to two-axion exchange between two neuclons (protons or neutrons)  can be described as \cite{V2014Constraints, 2003Constraining}

\begin{equation}
V(r)	=-\frac{g_{ak}^2  g_{al}^2}{32\pi^3 m^2} \frac{m_a}{r^2} K_1(2m_a r),
\end{equation}
provided that $r\gg 1/m$.
Here $g_{ak}$ and $g_{al}$ are the constants of a pseudoscalar axion-proton ($k, l=p$) or axion-neutron ($k, l=n$) interaction, $m=(m_n+m_p)/2$ is the mean of the neutron and proton masses,  $m_a$ is the mass of the axion,
$K_1(x)$ is the modified Bessel function, and $r$ is the distance between two neucleons. Then taking into account that the characteristic size of the nanosphere is several orders smaller than the alternative sections, and following \cite{V2014Constraints}, we obtain
\begin{equation}
\frac{\partial F_{add}^\beta (d)}{\partial d}=\frac{\pi}{m^2 m_{H}^2} C_{\beta} C_s I,
\end{equation}
with
\begin{align}
 I= \int_{1}^{\infty}du \frac{\sqrt { u^2-1}}{u^2}(1-e^{-2m_a u D})\notag \\ \times e^{-2m_a du}\Phi(R, m_a u) ,
\end{align}

where the following notation is introduced:
\begin{equation}
\Phi(r,z)=r-\frac{1}{2z}+e^{-2rz}(r+\frac{1}{2z}).
\end{equation}
Here $\beta=Au, Al$, $m_H$ is the mass of the atomic hydrogen, and the coefficients $C_{\beta, s}$ are defined as
\begin{equation}
	C_{\beta, s} = \rho_{\beta,s} (\frac{g_{ap}^2}{4\pi} \frac{Z_{\beta, s}}{\mu_{\beta, s}}+\frac{g_{an}^2}{4\pi} \frac{N_{\beta, s}}{\mu_{\beta, s}}), 
\end{equation}
where $\rho_{\beta,s}$ is the densities of Au(Al) and the sphere, $Z_{\beta, s}$ and $N_{\beta, s}$ are the number of protons and the mean number of neutrons in the atom Au(Al) and the molecule $SiO_2$, and the quantities $\mu_{\beta, s}$ are defined as $\mu_{\beta, s}=m_{\beta, s}/m_H$ where 
$m_{\beta, s}$ is the mean mass of the atom Au (Al)
and the molecule $SiO_2$. Moreover, from Eq. (25) we derive
\begin{equation}
	\frac{\partial F_{add}^{Al} (d)}{\partial d}
	-\frac{\partial F_{add}^{Au} (d)}{\partial d}
	=\frac{\pi}{m^2 m_{H}^2} (C_{Al}-C_{Au}) C_s I.
\end{equation}
Note that in Eqs. (24)-(29) the system of natural units is used.
Till now, we have established the relationship between the differences of the  modified resonance frequencies  $\omega^{\prime}_{Au} - \omega^{\prime}_{Al}$ and the unknown constants $g_{ap}$ and $g_{an}$ (Eq. (23) and Eqs.(26)-(29)). Next we demonstrate how to determine  $\omega^{\prime}_{Au} - \omega^{\prime}_{Al}$. 
\begin{figure*}
	\includegraphics[width=50em]{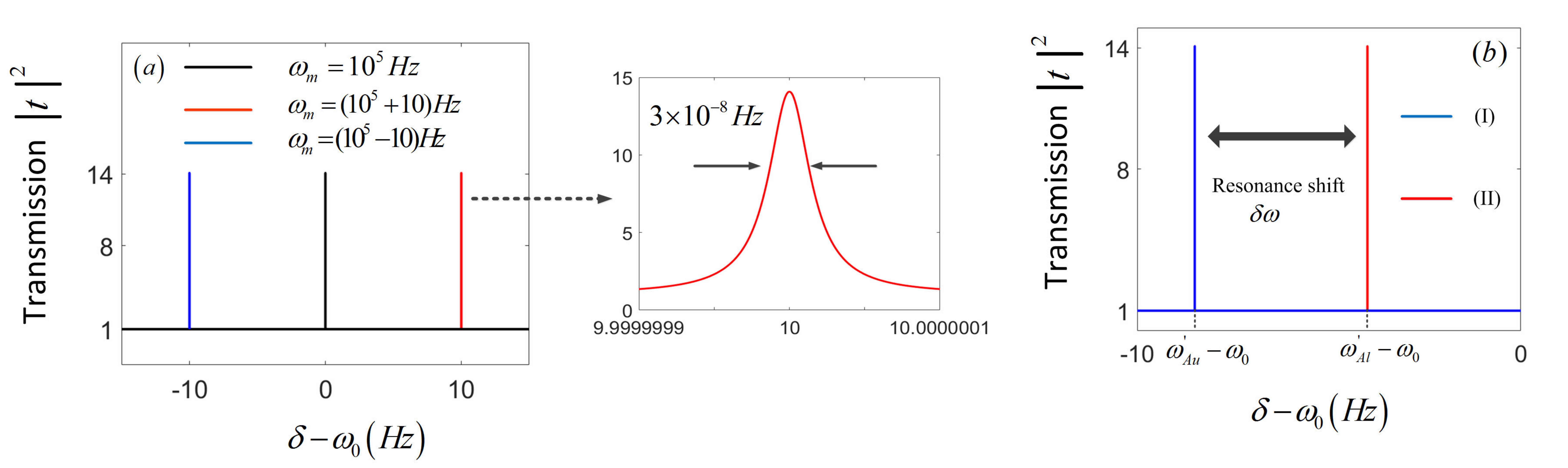}
	\caption{\label{fig:epsart} (a) We plot  the transmission ($|t|^2$) as a function of $\delta-\omega_0$ when $\omega_m$ take different values. The black, red, blue curves refer to the cases of $\omega_m =10^{5}, (10^{5}+10), (10^{5}-10)Hz$ respectively. The parameters used are  $\omega_0 \sim 100kHz$, $\kappa =1MHz$, $\gamma_m\approx 3.33\times 10^{-8}Hz$, $\Delta=0$,  $g=200Hz$, $E_{pu}=1kHz$, $E_{pr}=100Hz$.
	The middle window is the enlarged red peak with a linewidth of $3\times10^{-8} Hz$. (b) The tranamission spectrums generated experimentally are simulated by the blue and the red curves, which correspond to the trapping positions of type (I) and type (II) respectively. The distance between two peaks is designated as $\delta \omega$.}
\end{figure*}

 In terms of our scheme where the trapping position of the nanosphere is translated along the surface with a constant separation, we choose feasible parameters to investigate the transmission. The actual resonance frequency of the nanosphere is either
$\omega_{Au}^{\prime}$ or $\omega_{Al}^{\prime}$ which depends on the specific trapping position, while the 
unperturbed one can be assumed as $\omega_0 \sim 100kHz$ \cite{2020Optomechanics}. The mechanical quality factor of the nanosphere can be selected as $Q\sim 3\times 10^{12}$ \cite{2009Cavity}. Thus the damping rate can be calculated as
$\gamma_m=\frac{\omega_0}Q \approx 3.33\times 10^{-8}Hz$. The decay rate of the cavity amplitude can be chosen as $\kappa =1MHz$ \cite{2009Observation}. For both types of  positions, the parameters used are $\Delta=0$,  $g=200Hz$, $E_{pu}=1kHz$, and $E_{pr}=100Hz$.

Provided that the parameters $(\omega_0, \kappa, \gamma_m, \Delta, g, E_{pr}, E_{pu} )$ take values as the above while  $\omega_m$ is specified as $\omega_{m}=10^5 Hz, (10^5\pm 10) Hz$ respectively,  according to Eqs. (14), (15) and (17), we plot the transmission ($|t|^2$) as a function of $\delta-\omega_0$ in Fig. 3(a). Let us focus on Fig. 3(a).  For the three transmission spectrums plotted with different colors, three resonance peaks the linewidth (full width at half maximum) of which are both $3\times 10^{-8}Hz$ appear at  
-10, 0, 10 Hz respectively, while the rest of the spectrums  concide with each other. We designate this linewidth  as $\Delta f$ in the following.
With more numerical analysis, we can conclude that the very transmission spectrum at $-100 Hz \le \delta-\omega_0\le 100 Hz$, plotted with $\omega_{m}$  satisfying  $\omega_0-100Hz\le \omega_m \le \omega_0+100Hz$ and other parameters remaining unchanged, is composed of a straight  horizontal line  with vertical coordinate 1  and a resonance peak at $\omega_m -\omega_0$, just like the three spectrums in Fig. 3(a).

 Suppose that our scheme  where the parameters $(\omega_0, \kappa, \gamma_m, \Delta, g, E_{pr}, E_{pu} )$ take vlaues as in Fig. 3(a) has been realized experimentally.  The corresponding generated transmission spectrums are represented by the plot in Fig. 3(b). Since the nanosphere is trapped in position (I) ((II)), the resonance frequency of it would be  $\omega_m=\omega_{Au}^\prime$ ($\omega_m=\omega_{Al}^\prime$), and  a resonance peak would appear at $\omega_{Au}^{\prime}-\omega_0$ ($\omega_{Al}^{\prime}-\omega_0$)in the  corresponding tranmission spectrum as shown by the blue and the red curves in (b).  Note that one straight red line, which is a part of the red curve, is overlapped by the horizontal blue line. In addition, it also should be noted that (b) does not imply $\omega_{Al}^{\prime}-\omega_0 >  \omega_{Au}^{\prime}-\omega_0$ since we do not know which one of $\omega_{Al}^{\prime}$ and $\omega_{Au}^{\prime}$ is relative bigger than the other.
 Then  with the translation of the trapping positions, we can observe that one peak appears at two locations ($\omega_{Al}^{\prime}-\omega_0$ , $\omega_{Au}^{\prime}-\omega_0$) alternatively, i.e. one peak shifts between two positions back and forth with a displacement
 \begin{align}
 \delta\omega&=|(\omega_{Al}^{\prime}-\omega_0 )-(\omega_{Au}^{\prime}-\omega_0 )|\notag\\&=|\omega_{Al}^{\prime}-\omega_{Au}^{\prime}|.
 \end{align}
 Till now, we have established a detection method via Eqs. (23), (29) and (30). Firstly, we measure resonace shift $\delta\omega$ in the relevant experiment. Secondly, substituting fitting values into the parameters in Eqs. (23), (29) and (30), combining these equations and performing calculation, we can determine the values of $g_{ap}$ or $g_{an}$ or set constraints on them provided that specific conditions (see the following) are satisfied. 
 \section{\label{sec:level1}Noise analysis and prospective constraints}
 Since the measureable quantity (i.e. resonance shift $\delta\omega$) has been connected with the unknown constants  $g_{ap}$ and $g_{an}$,  to set upbounds on these constants via minimum detectable resonance shift $\delta\omega_{min}$ is natural. Then the question is to obtain a resonable value of $\delta\omega_{min}$, which is resolved by two different approaches in the following.

 Firstly, we derive the fundamental limit imposed by noise. Since nano-particles levitated in optical fields act as nanoscale oscillators \cite{2020Optomechanics}, the methods of noise analysis developed in nanomechanical systems \cite{2002Noise, Ekinci2004Ultimate} can be ultilized here.  In our proposed system, the thermal noise, which originates from thermally driven random motion of the mechanical device, is the dominant noise source. According to \cite{Ekinci2004Ultimate}, the minimum measurable frequency shift $\delta \omega_m$ can be expressed as
 \begin{equation}
 	\delta \omega_m=\left[\frac{K_B T}{E_C}\frac{\omega_0 \Delta f}{Q} \right]^{1/2},
 \end{equation}
 where $E_C=M_{eff} \omega_0^2 \langle x_c^2 \rangle $, $M_{eff}$ is the effcctive mass of the nanosphere, $\langle x_c \rangle $ is the constant mean square amplitude of it which is driven in a mesurement, $K_B$ is Boltamann's constant and $T$ is the temprature of the nanosphere.  Because in our scheme the frequency shift is equal to the resonance shift (see Eq. (30)), $\delta \omega_m$ in Eq. (31) can be subsituted with
 $\delta\omega_{min}$. Let the parameters in Eq. (31)
 take appropriate values as follows: $M_{eff}=m_s=1.05\times 10^{-20}kg$, $\omega_0=10^5 Hz$, 
 $\Delta f=3\times 10^{-8}Hz$, $K_B=1.38\times 10^{-23} J\bullet K^{-1}$, $Q=3\times 10^{12}$, $ T$ and $\langle x^2 \rangle$ can be assumed as $T=1mk$ \cite{2020Optomechanics} and $\langle x^2 \rangle= 100 nm^2$ \cite{Ekinci2004Ultimate} respectively. Substituting these values into Eq. (31) and $\delta \omega_m$ with  $\delta\omega_{min}$, we derive $\delta\omega_{min}= 3.6253\times 10^{-8}Hz$.

 Secondly, we investigate the minimum detectable resonance shift from an experimental viewpoint. In the frequency detection regime, the detection limit is closely related to the relavant linewidth \cite{2018Probing, Ekinci2004Ultimate}. Moreover,
 F. Vollmer et al.\cite{2012Review} present an example where the binding of molecules on the surface of a resonator shifts the resonant frequency (see Figure 8 in \cite{2012Review} ). The demonstration relating to the example (see the caption of this figure) implies that the linewidth can be assumed as the minimum detectable resonance shift. Then we assume
  $\delta\omega_{min}= \Delta f=3\times10^{-8}Hz$. Now we see that  the values of $\delta\omega_{min}$ obtained from the above two approaches are consistent. Due to this, we use the linewidth as the minimum detectable resonance shift to set  prospective constraints in the following.

 Suppose in the relavant experiment we may not observe a  resonance shift, i.e. 
 \begin{equation}
 \delta\omega<\delta\omega_{min}=3\times10^{-8}Hz.
 \end{equation}
 By Eqs. (23), (30) and (32), we can derive that
 \begin{equation}
\small |\frac{\partial F_{add}^{Al} (d)}{\partial d}
 -\frac{\partial F_{add}^{Au} (d)}{\partial d}|<2m_s \omega_0 \delta\omega_{min}=6.2832\times10^{-23}kg/s^2.
 \end{equation}
 As $1kg/s^2=2.4313\times10^5(eV)^3/\hbar^2 c^2$, we further obtain
  \begin{equation}
 	\small |\frac{\partial F_{add}^{Al} (d)}{\partial d}
 	-\frac{\partial F_{add}^{Au} (d)}{\partial d}|<1.5276\times10^{-17}(eV)^3/\hbar^2 c^2.
 \end{equation}
 Substituting Eq. (34) into Eq. (29) where the system of natural units is used, we derive 
 \begin{equation}
 |\frac{\pi}{m^2 m_{H}^2} (C_{Al}-C_{Au}) C_s I|<1.5276\times10^{-17}(eV)^3.
 \end{equation}
 Using Eqs. (28) and (35), under the conditions of $g_{ap}^2\gg g_{an}^2$,  $g_{an}^2\gg g_{ap}^2$ and
 $g_{an}^2= g_{ap}^2$ respectively, we can derive three inequalities as follows. For $g_{ap}^2\gg g_{an}^2$, we obtain
  \begin{equation}
 	\frac{g_{ap}^2}{4\pi}<m m_H\sqrt{\frac{1.5276\times 10^{-17}(eV)^3}{|\pi(\rho_{Al}\frac{Z_{Al}}{\mu_{Al}}-\rho_{Au}\frac{Z_{Au}}{\mu_{Au}})\rho_{s}\frac{Z_{s}}{\mu_{s}}I|}},
 \end{equation}
for $g_{an}^2\gg g_{ap}^2$,
  \begin{equation}
 	\frac{g_{an}^2}{4\pi}<m m_H\sqrt{\frac{1.5276\times 10^{-17}(eV)^3}{|\pi(\rho_{Al}\frac{N_{Al}}{\mu_{Al}}-\rho_{Au}\frac{N_{Au}}{\mu_{Au}})\rho_{s}\frac{N_{s}}{\mu_{s}}I|}},
 \end{equation}
 and for $g_{an}^2= g_{ap}^2$,
\begin{widetext}
	\begin{equation}
		\frac{g_{an}^2}{4\pi}<m m_H\sqrt{\frac{1.5276\times 10^{-17}(eV)^3}{|\pi\left[\rho_{Al}(\frac{Z_{Al}}{\mu_{Al}}+\frac{N_{Al}}{\mu_{Al}})-\rho_{Au}(\frac{Z_{Au}}{\mu_{Au}}+\frac{N_{Au}}{\mu_{Au}})\right]
				\rho_{s}(\frac{Z_{s}}{\mu_{s}}+\frac{N_{s}}{\mu_{s}})I|}}.	
	\end{equation}
\end{widetext}
 Note that in Eqs. (35)-(38) the system of natural units is used. The values of several parameters in these equations can be found in \cite{V2014Constraints}:
 \begin{align}
 	\frac{Z_{Al}}{\mu_{Al}}&=0.48558, & 	\frac{N_{Al}}{\mu_{Al}}&=0.52304,\notag\\
 \frac{Z_{Au}}{\mu_{Au}}&=0.40422, & 	\frac{N_{Au}}{\mu_{Au}}&=0.60378,\notag\\
 \frac{Z_{s}}{\mu_{s}}&=0.503205, & 	\frac{N_{s}}{\mu_{s}}&=0.505179.
\end{align}
Also, three densities in natural units are calculated as:
\begin{align}
	\rho_{Al}=1.2\times10^{-5} (MeV)^4,\notag\\
	\rho_{Au}=8.3\times10^{-5} (MeV)^4,\notag\\
	\rho_{s}=1.1\times10^{-5} (MeV)^4.
\end{align}
And, the values of $m$ and $m_H$ are obtained:
\begin{align}
m=938.9150MeV, \notag\\m_H=938.771MeV.
\end{align}

\begin{figure}
	\includegraphics[width=28em]{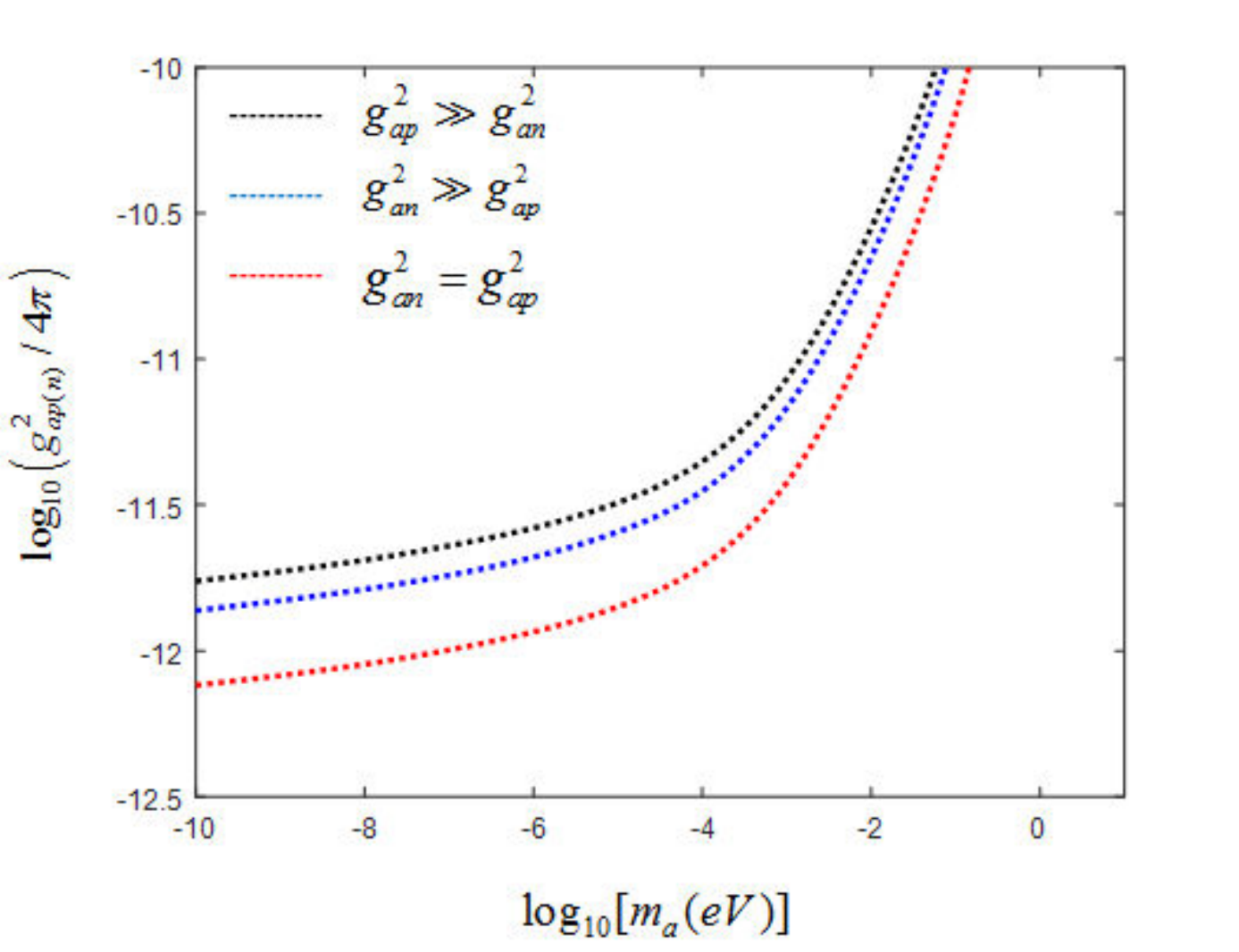}
	\caption{\label{fig:epsart} Constraints on  the constant of axion-proton (neutron) interaction as functions of the axion mass $m_a$. The black, blue, and red lines are obtained under the conditions $g_{ap}^2\gg g_{an}^2$, $g_{an}^2\gg g_{ap}^2$, and $g_{an}^2= g_{ap}^2$ respectively. }
\end{figure}
Substituting  Eqs. (39)-(41) and the values of $D$, $R$ and $d$ in natural units into Eqs. (36)-(38) and then performing computation, we attain three upbounds for $g_{ap(n)}^2/4\pi$ as functions of $m_a$ as shown in Fig. 4. Note that these upbounds  correspond to $g_{ap}^2\gg g_{an}^2$,  $g_{an}^2\gg g_{ap}^2$, and $g_{an}^2= g_{ap}^2$ respectively.

we compare our obtained constraints under the most reasonable condition $g_{an}^2= g_{ap}^2$ \cite{2003Constraining} with existing  strongest laboratory constraints on axion-like particles via Fig. 5. The upper limit  at $m_a<1\mu eV$ was obtained with a magnetometer \cite{2009Limits} (line 1). The constraints  most stringent at $1\mu eV< m_a <1.7meV$ were established in \cite{2006Particle}  by ultilizing the data from the search for violations of the gravitational inverse-square law \cite{2007Tests1} (line 2). The upper limit at about $1meV < m_a <0.5eV$  was derived in \cite{2015Improved} from the measurement results of the Casimir-less experiment \cite{2016Stronger} (line 3). The most stringent constraints at $m_a>0.5eV$ were obtained from experiments on measuring the forces between protons in the beam of molecular hydrogen \cite{1979The, 2013Constraints} (line 4). Our work is represented by the red curve 5, which is consistent with the bottom line in Fig. 4. As can be seen in the Fig. 5, our constraints significantly improve the upper limit in the wide range of axion mass approximately from $10^{-4} \mu eV$ to $10$ $ eV$. 
\begin{figure}
	\includegraphics[width=28em]{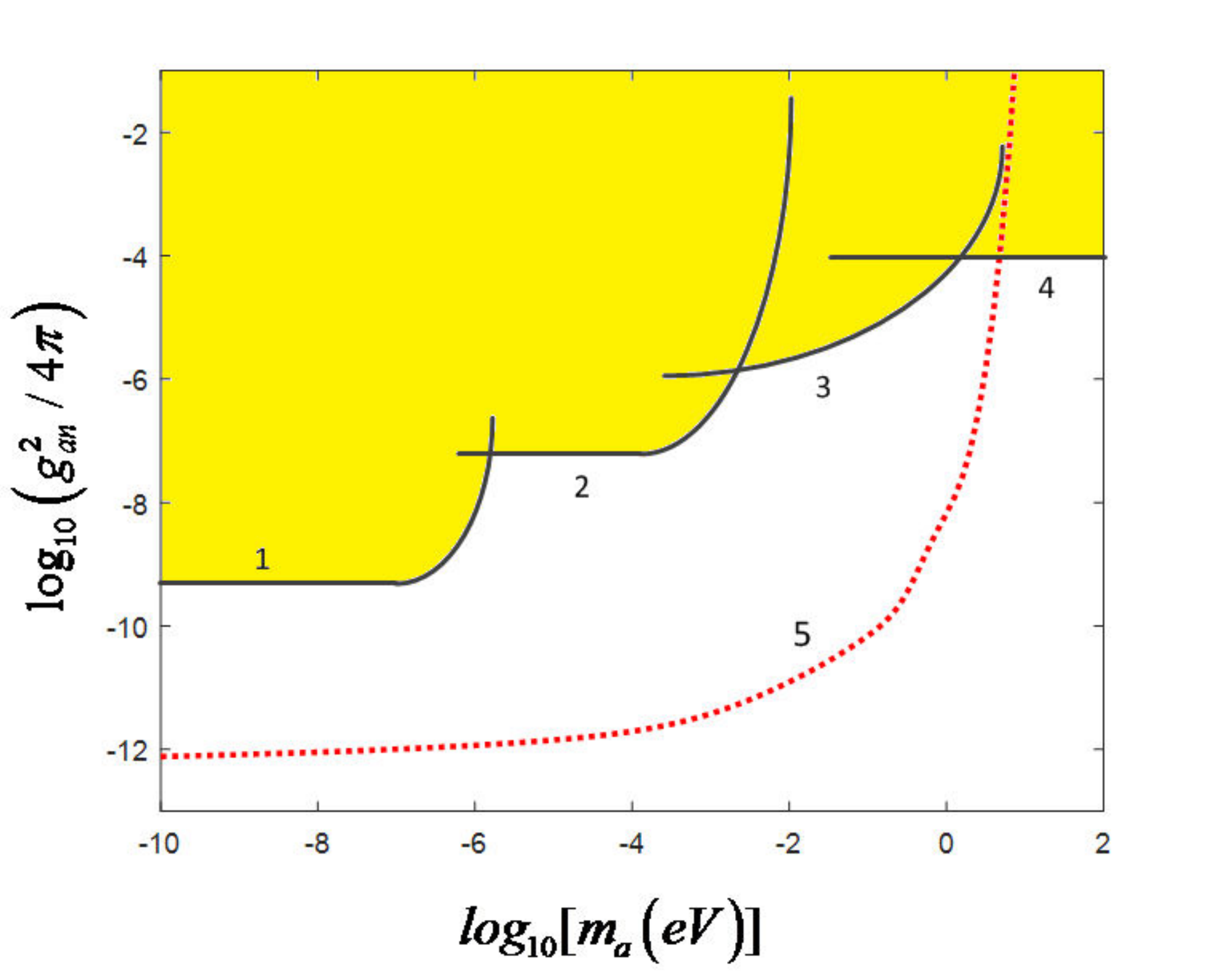}
	\caption{\label{fig:epsart} Constraints on  the constant of axion-neutron interaction under the condition $g_{an}^2= g_{ap}^2$ from the measurement of changes in the precession frequency \cite{2009Limits} (line 1), from the search for violations of the gravitational inverse-square law \cite{2006Particle,2007Tests1} (line 2),  from a Casimir-less experiment \cite{2015Improved, 2016Stronger} (line 3) , from measuring the forces between protons \cite{1979The, 2013Constraints} (line 4), and from our work (red line). The yellow region is excluded.
	}
\end{figure}

\section{\label{sec:level1}conclusion and discussion}
In sum, we develop an optical method to constrain the axion-neucleon interaction. When we tanslate the trapping positions of the nanosphere, the actual resonance frequency of it changes. Via substraction, this frequency shift can be ralate to the additional forces acted on the nanosphere . On the other hand, this shift  can be determined from measuring the resonance shift in the transmission spectrum. Then based on the estimation of the minimum detectable resonance shift, we derive the prospective constraints in the case of $g_{an}^2=g_{ap}^2$, which improve on the previous bounds by several orders of magnitude at an ultrawide axion mass range .

Recently, experimental searches for the Yukawa interaction \cite{2005Constraining,2016Stronger} or testing Newtonian gravity \cite{2003Testing} in the Casimir-less regime have been conducted, and often constraints on non-Newtonian gravity and axion-like particles were derived from the results of the same experiments \cite{PhysRevD.101.056013, 2014Constraints,2010Advance, 2016Stronger, 2015Improved}. Besides, for the purpose of detecting non-Newtonain gravity,  Geraci et al. \cite{ 2010Short} propose a Casimir-less regime where optically cooled levitated microspheres are ultilized. All these imply that the application of our method can be extended to the areas of constraining Yukawa interaction and testing Newtonian gravity. Finally, we expect our work can be realized experimentally in the near future.

\begin{acknowledgments}

National Natural Science Foundation of China (Nos.11274230 and 11574206) and Natural Science Foundation of Shanghai (No.20ZR1429900).
\end{acknowledgments}

\bibliography{apssamp}

\begin{thebibliography}{52}%
\makeatletter
\providecommand \@ifxundefined [1]{%
 \@ifx{#1\undefined}
}%
\providecommand \@ifnum [1]{%
 \ifnum #1\expandafter \@firstoftwo
 \else \expandafter \@secondoftwo
 \fi
}%
\providecommand \@ifx [1]{%
 \ifx #1\expandafter \@firstoftwo
 \else \expandafter \@secondoftwo
 \fi
}%
\providecommand \natexlab [1]{#1}%
\providecommand \enquote  [1]{``#1''}%
\providecommand \bibnamefont  [1]{#1}%
\providecommand \bibfnamefont [1]{#1}%
\providecommand \citenamefont [1]{#1}%
\providecommand \href@noop [0]{\@secondoftwo}%
\providecommand \href [0]{\begingroup \@sanitize@url \@href}%
\providecommand \@href[1]{\@@startlink{#1}\@@href}%
\providecommand \@@href[1]{\endgroup#1\@@endlink}%
\providecommand \@sanitize@url [0]{\catcode `\\12\catcode `\$12\catcode
  `\&12\catcode `\#12\catcode `\^12\catcode `\_12\catcode `\%12\relax}%
\providecommand \@@startlink[1]{}%
\providecommand \@@endlink[0]{}%
\providecommand \url  [0]{\begingroup\@sanitize@url \@url }%
\providecommand \@url [1]{\endgroup\@href {#1}{\urlprefix }}%
\providecommand \urlprefix  [0]{URL }%
\providecommand \Eprint [0]{\href }%
\providecommand \doibase [0]{https://doi.org/}%
\providecommand \selectlanguage [0]{\@gobble}%
\providecommand \bibinfo  [0]{\@secondoftwo}%
\providecommand \bibfield  [0]{\@secondoftwo}%
\providecommand \translation [1]{[#1]}%
\providecommand \BibitemOpen [0]{}%
\providecommand \bibitemStop [0]{}%
\providecommand \bibitemNoStop [0]{.\EOS\space}%
\providecommand \EOS [0]{\spacefactor3000\relax}%
\providecommand \BibitemShut  [1]{\csname bibitem#1\endcsname}%
\let\auto@bib@innerbib\@empty
\bibitem [{\citenamefont {Weinberg}(1978)}]{PhysRevLett.40.223}%
  \BibitemOpen
  \bibfield  {author} {\bibinfo {author} {\bibfnamefont {S.}~\bibnamefont
  {Weinberg}},\ }\bibfield  {title} {\bibinfo {title} {A new light boson?},\
  }\href {https://doi.org/10.1103/PhysRevLett.40.223} {\bibfield  {journal}
  {\bibinfo  {journal} {Phys. Rev. Lett.}\ }\textbf {\bibinfo {volume} {40}},\
  \bibinfo {pages} {223} (\bibinfo {year} {1978})}\BibitemShut {NoStop}%
\bibitem [{\citenamefont {Wilczek}(1978)}]{PhysRevLett.40.279}%
  \BibitemOpen
  \bibfield  {author} {\bibinfo {author} {\bibfnamefont {F.}~\bibnamefont
  {Wilczek}},\ }\bibfield  {title} {\bibinfo {title} {Problem of strong $p$ and
  $t$ invariance in the presence of instantons},\ }\href
  {https://doi.org/10.1103/PhysRevLett.40.279} {\bibfield  {journal} {\bibinfo
  {journal} {Phys. Rev. Lett.}\ }\textbf {\bibinfo {volume} {40}},\ \bibinfo
  {pages} {279} (\bibinfo {year} {1978})}\BibitemShut {NoStop}%
\bibitem [{\citenamefont {Ringwald}(2014)}]{2014Experimental}%
  \BibitemOpen
  \bibfield  {author} {\bibinfo {author} {\bibfnamefont {A.}~\bibnamefont
  {Ringwald}},\ }\bibfield  {title} {\bibinfo {title} {Experimental searches
  for axions and axion-like particles},\ }\href@noop {} {\bibfield  {journal}
  {\bibinfo  {journal} {Archivos Espaoles De Urología}\ }\textbf {\bibinfo
  {volume} {60}},\ \bibinfo {pages} {815} (\bibinfo {year} {2014})}\BibitemShut
  {NoStop}%
\bibitem [{\citenamefont {Beringer}\ \emph {et~al.}(2012)\citenamefont
  {Beringer}, \citenamefont {Arguin}, \citenamefont {Barnett}, \citenamefont
  {Copic}, \citenamefont {Dahl}, \citenamefont {Groom}, \citenamefont {Lin},
  \citenamefont {Lys}, \citenamefont {Murayama}, \citenamefont {Wohl},
  \citenamefont {Yao}, \citenamefont {Zyla}, \citenamefont {Amsler},
  \citenamefont {Antonelli}, \citenamefont {Asner}, \citenamefont {Baer},
  \citenamefont {Band}, \citenamefont {Basaglia}, \citenamefont {Bauer},\ and\
  \citenamefont {Beatty}}]{PhysRevD.86.010001}%
  \BibitemOpen
  \bibfield  {author} {\bibinfo {author} {\bibfnamefont {J.}~\bibnamefont
  {Beringer}}, \bibinfo {author} {\bibfnamefont {J.~F.}\ \bibnamefont
  {Arguin}}, \bibinfo {author} {\bibfnamefont {R.~M.}\ \bibnamefont {Barnett}},
  \bibinfo {author} {\bibfnamefont {K.}~\bibnamefont {Copic}}, \bibinfo
  {author} {\bibfnamefont {O.}~\bibnamefont {Dahl}}, \bibinfo {author}
  {\bibfnamefont {D.~E.}\ \bibnamefont {Groom}}, \bibinfo {author}
  {\bibfnamefont {C.~J.}\ \bibnamefont {Lin}}, \bibinfo {author} {\bibfnamefont
  {J.}~\bibnamefont {Lys}}, \bibinfo {author} {\bibfnamefont {H.}~\bibnamefont
  {Murayama}}, \bibinfo {author} {\bibfnamefont {C.~G.}\ \bibnamefont {Wohl}},
  \bibinfo {author} {\bibfnamefont {W.~M.}\ \bibnamefont {Yao}}, \bibinfo
  {author} {\bibfnamefont {P.~A.}\ \bibnamefont {Zyla}}, \bibinfo {author}
  {\bibfnamefont {C.}~\bibnamefont {Amsler}}, \bibinfo {author} {\bibfnamefont
  {M.}~\bibnamefont {Antonelli}}, \bibinfo {author} {\bibfnamefont {D.~M.}\
  \bibnamefont {Asner}}, \bibinfo {author} {\bibfnamefont {H.}~\bibnamefont
  {Baer}}, \bibinfo {author} {\bibfnamefont {H.~R.}\ \bibnamefont {Band}},
  \bibinfo {author} {\bibfnamefont {T.}~\bibnamefont {Basaglia}}, \bibinfo
  {author} {\bibfnamefont {C.~W.}\ \bibnamefont {Bauer}},\ and\ \bibinfo
  {author} {\bibfnamefont {J.~J.}\ \bibnamefont {Beatty}} (\bibinfo
  {collaboration} {Particle Data Group}),\ }\bibfield  {title} {\bibinfo
  {title} {Review of particle physics},\ }\href
  {https://doi.org/10.1103/PhysRevD.86.010001} {\bibfield  {journal} {\bibinfo
  {journal} {Phys. Rev. D}\ }\textbf {\bibinfo {volume} {86}},\ \bibinfo
  {pages} {010001} (\bibinfo {year} {2012})}\BibitemShut {NoStop}%
\bibitem [{\citenamefont {Tanabashi}\ \emph {et~al.}(2018)\citenamefont
  {Tanabashi}, \citenamefont {Hagiwara}, \citenamefont {Hikasa}, \citenamefont
  {Nakamura}, \citenamefont {Sumino}, \citenamefont {Takahashi}, \citenamefont
  {Tanaka}, \citenamefont {Agashe}, \citenamefont {Aielli}, \citenamefont
  {Amsler}, \citenamefont {Antonelli}, \citenamefont {Asner}, \citenamefont
  {Baer}, \citenamefont {Banerjee}, \citenamefont {Barnett}, \citenamefont
  {Basaglia}, \citenamefont {Bauer},\ and\ \citenamefont
  {Beatty}}]{PhysRevD.98.030001}%
  \BibitemOpen
  \bibfield  {author} {\bibinfo {author} {\bibfnamefont {M.}~\bibnamefont
  {Tanabashi}}, \bibinfo {author} {\bibfnamefont {K.}~\bibnamefont {Hagiwara}},
  \bibinfo {author} {\bibfnamefont {K.}~\bibnamefont {Hikasa}}, \bibinfo
  {author} {\bibfnamefont {K.}~\bibnamefont {Nakamura}}, \bibinfo {author}
  {\bibfnamefont {Y.}~\bibnamefont {Sumino}}, \bibinfo {author} {\bibfnamefont
  {F.}~\bibnamefont {Takahashi}}, \bibinfo {author} {\bibfnamefont
  {J.}~\bibnamefont {Tanaka}}, \bibinfo {author} {\bibfnamefont
  {K.}~\bibnamefont {Agashe}}, \bibinfo {author} {\bibfnamefont
  {G.}~\bibnamefont {Aielli}}, \bibinfo {author} {\bibfnamefont
  {C.}~\bibnamefont {Amsler}}, \bibinfo {author} {\bibfnamefont
  {M.}~\bibnamefont {Antonelli}}, \bibinfo {author} {\bibfnamefont {D.~M.}\
  \bibnamefont {Asner}}, \bibinfo {author} {\bibfnamefont {H.}~\bibnamefont
  {Baer}}, \bibinfo {author} {\bibfnamefont {S.}~\bibnamefont {Banerjee}},
  \bibinfo {author} {\bibfnamefont {R.~M.}\ \bibnamefont {Barnett}}, \bibinfo
  {author} {\bibfnamefont {T.}~\bibnamefont {Basaglia}}, \bibinfo {author}
  {\bibfnamefont {C.~W.}\ \bibnamefont {Bauer}},\ and\ \bibinfo {author}
  {\bibfnamefont {J.~J.}\ \bibnamefont {Beatty}} (\bibinfo {collaboration}
  {Particle Data Group}),\ }\bibfield  {title} {\bibinfo {title} {Review of
  particle physics},\ }\href {https://doi.org/10.1103/PhysRevD.98.030001}
  {\bibfield  {journal} {\bibinfo  {journal} {Phys. Rev. D}\ }\textbf {\bibinfo
  {volume} {98}},\ \bibinfo {pages} {030001} (\bibinfo {year}
  {2018})}\BibitemShut {NoStop}%
\bibitem [{\citenamefont {Masahiro}\ \emph {et~al.}(2013)\citenamefont
  {Masahiro}, \citenamefont {Kawasaki}, \citenamefont {Kazunori},\ and\
  \citenamefont {Nakayama}}]{Masahiro2013Axions}%
  \BibitemOpen
  \bibfield  {author} {\bibinfo {author} {\bibnamefont {Masahiro}}, \bibinfo
  {author} {\bibnamefont {Kawasaki}}, \bibinfo {author} {\bibnamefont
  {Kazunori}},\ and\ \bibinfo {author} {\bibnamefont {Nakayama}},\ }\bibfield
  {title} {\bibinfo {title} {Axions: Theory and cosmological role},\
  }\href@noop {} {\bibfield  {journal} {\bibinfo  {journal} {Annual Review of
  Nuclear and Particle Science}\ }\textbf {\bibinfo {volume} {63}} (\bibinfo
  {year} {2013})}\BibitemShut {NoStop}%
\bibitem [{\citenamefont {Irastorza}\ and\ \citenamefont
  {Javier}(2018)}]{2018New}%
  \BibitemOpen
  \bibfield  {author} {\bibinfo {author} {\bibfnamefont {I.~G.}\ \bibnamefont
  {Irastorza}}\ and\ \bibinfo {author} {\bibfnamefont {R.}~\bibnamefont
  {Javier}},\ }\bibfield  {title} {\bibinfo {title} {New experimental
  approaches in the search for axion-like particles},\ }\href@noop {}
  {\bibfield  {journal} {\bibinfo  {journal} {Progress in Particle Nuclear
  Physics}\ ,\ \bibinfo {pages} {S014664101830036X}} (\bibinfo {year}
  {2018})}\BibitemShut {NoStop}%
\bibitem [{\citenamefont {Ficek}\ and\ \citenamefont
  {Budker}(2018)}]{2018Constraining}%
  \BibitemOpen
  \bibfield  {author} {\bibinfo {author} {\bibfnamefont {F.}~\bibnamefont
  {Ficek}}\ and\ \bibinfo {author} {\bibfnamefont {D.}~\bibnamefont {Budker}},\
  }\bibfield  {title} {\bibinfo {title} {Constraining exotic interactions},\
  }\href@noop {} {\bibfield  {journal} {\bibinfo  {journal} {Annalen Der
  Physik}\ } (\bibinfo {year} {2018})}\BibitemShut {NoStop}%
\bibitem [{\citenamefont {Safronova}\ \emph {et~al.}(2018)\citenamefont
  {Safronova}, \citenamefont {Budker}, \citenamefont {Demille}, \citenamefont
  {Kimball}, \citenamefont {Derevianko},\ and\ \citenamefont
  {Clark}}]{2017Search}%
  \BibitemOpen
  \bibfield  {author} {\bibinfo {author} {\bibfnamefont {M.~S.}\ \bibnamefont
  {Safronova}}, \bibinfo {author} {\bibfnamefont {D.}~\bibnamefont {Budker}},
  \bibinfo {author} {\bibfnamefont {D.}~\bibnamefont {Demille}}, \bibinfo
  {author} {\bibfnamefont {D.}~\bibnamefont {Kimball}}, \bibinfo {author}
  {\bibfnamefont {A.}~\bibnamefont {Derevianko}},\ and\ \bibinfo {author}
  {\bibfnamefont {C.~W.}\ \bibnamefont {Clark}},\ }\bibfield  {title} {\bibinfo
  {title} {Search for new physics with atoms and molecules},\ }\href@noop {}
  {\bibfield  {journal} {\bibinfo  {journal} {Rev. Mod. Phys}\ }\textbf
  {\bibinfo {volume} {90}},\ \bibinfo {pages} {025008} (\bibinfo {year}
  {2018})}\BibitemShut {NoStop}%
\bibitem [{\citenamefont {Rosenberg}\ and\ \citenamefont {van
  Bibber}(2000)}]{Leslie2000Searches}%
  \BibitemOpen
  \bibfield  {author} {\bibinfo {author} {\bibfnamefont {L.~J.}\ \bibnamefont
  {Rosenberg}}\ and\ \bibinfo {author} {\bibfnamefont {K.~A.}\ \bibnamefont
  {van Bibber}},\ }\bibfield  {title} {\bibinfo {title} {Searches for invisible
  axions},\ }\href@noop {} {\bibfield  {journal} {\bibinfo  {journal} {Physics
  Reports}\ } (\bibinfo {year} {2000})}\BibitemShut {NoStop}%
\bibitem [{\citenamefont {Braine}\ \emph {et~al.}(2020)\citenamefont {Braine},
  \citenamefont {Cervantes}, \citenamefont {Crisosto}, \citenamefont {Du},\
  and\ \citenamefont {Murch}}]{2020Extended}%
  \BibitemOpen
  \bibfield  {author} {\bibinfo {author} {\bibfnamefont {T.}~\bibnamefont
  {Braine}}, \bibinfo {author} {\bibfnamefont {R.}~\bibnamefont {Cervantes}},
  \bibinfo {author} {\bibfnamefont {N.}~\bibnamefont {Crisosto}}, \bibinfo
  {author} {\bibfnamefont {N.}~\bibnamefont {Du}},\ and\ \bibinfo {author}
  {\bibfnamefont {K.~W.}\ \bibnamefont {Murch}},\ }\bibfield  {title} {\bibinfo
  {title} {Extended search for the invisible axion with the axion dark matter
  experiment},\ }\href@noop {} {\bibfield  {journal} {\bibinfo  {journal}
  {Phys. Rev. Lett}\ }\textbf {\bibinfo {volume} {124}},\ \bibinfo {pages}
  {101303} (\bibinfo {year} {2020})}\BibitemShut {NoStop}%
\bibitem [{\citenamefont {Sikivie}(2021)}]{2021Invisible}%
  \BibitemOpen
  \bibfield  {author} {\bibinfo {author} {\bibfnamefont {P.}~\bibnamefont
  {Sikivie}},\ }\bibfield  {title} {\bibinfo {title} {Invisible axion search
  methods},\ }\href@noop {} {\bibfield  {journal} {\bibinfo  {journal} {Rev.
  Mod. Phys}\ }\textbf {\bibinfo {volume} {93}},\ \bibinfo {pages} {015004}
  (\bibinfo {year} {2021})}\BibitemShut {NoStop}%
\bibitem [{\citenamefont {Klimchitskaya}(2017)}]{2017Recent}%
  \BibitemOpen
  \bibfield  {author} {\bibinfo {author} {\bibfnamefont {G.~L.}\ \bibnamefont
  {Klimchitskaya}},\ }\bibfield  {title} {\bibinfo {title} {Recent breakthrough
  and outlook in constraining the non-newtonian gravity and axion-like
  particles from casimir physics},\ }\href@noop {} {\bibfield  {journal}
  {\bibinfo  {journal} {Eur.Phys.J. C}\ }\textbf {\bibinfo {volume} {77}}
  (\bibinfo {year} {2017})}\BibitemShut {NoStop}%
\bibitem [{\citenamefont {Mostepanenko}\ and\ \citenamefont
  {Klimchitskaya}(2020)}]{2020The}%
  \BibitemOpen
  \bibfield  {author} {\bibinfo {author} {\bibfnamefont {V.~M.}\ \bibnamefont
  {Mostepanenko}}\ and\ \bibinfo {author} {\bibfnamefont {G.~L.}\ \bibnamefont
  {Klimchitskaya}},\ }\bibfield  {title} {\bibinfo {title} {The state of the
  art in constraining axion-to-nucleon coupling and non-newtonian gravity from
  laboratory experiments},\ }\href@noop {} {\bibfield  {journal} {\bibinfo
  {journal} {Universe}\ } (\bibinfo {year} {2020})}\BibitemShut {NoStop}%
\bibitem [{\citenamefont {Klimchitskaya}\ \emph {et~al.}(2020)\citenamefont
  {Klimchitskaya}, \citenamefont {Kuusk},\ and\ \citenamefont
  {Mostepanenko}}]{PhysRevD.101.056013}%
  \BibitemOpen
  \bibfield  {author} {\bibinfo {author} {\bibfnamefont {G.~L.}\ \bibnamefont
  {Klimchitskaya}}, \bibinfo {author} {\bibfnamefont {P.}~\bibnamefont
  {Kuusk}},\ and\ \bibinfo {author} {\bibfnamefont {V.~M.}\ \bibnamefont
  {Mostepanenko}},\ }\bibfield  {title} {\bibinfo {title} {Constraints on
  non-newtonian gravity and axionlike particles from measuring the casimir
  force in nanometer separation range},\ }\href
  {https://doi.org/10.1103/PhysRevD.101.056013} {\bibfield  {journal} {\bibinfo
   {journal} {Phys. Rev. D}\ }\textbf {\bibinfo {volume} {101}},\ \bibinfo
  {pages} {056013} (\bibinfo {year} {2020})}\BibitemShut {NoStop}%
\bibitem [{\citenamefont {Vasilakis}\ \emph {et~al.}(2009)\citenamefont
  {Vasilakis}, \citenamefont {Brown}, \citenamefont {Kornack},\ and\
  \citenamefont {Romalis}}]{2009Limits}%
  \BibitemOpen
  \bibfield  {author} {\bibinfo {author} {\bibfnamefont {G.}~\bibnamefont
  {Vasilakis}}, \bibinfo {author} {\bibfnamefont {J.~M.}\ \bibnamefont
  {Brown}}, \bibinfo {author} {\bibfnamefont {T.~W.}\ \bibnamefont {Kornack}},\
  and\ \bibinfo {author} {\bibfnamefont {M.~V.}\ \bibnamefont {Romalis}},\
  }\bibfield  {title} {\bibinfo {title} {Limits on new long range nuclear
  spin-dependent forces set with a k-3he co-magnetometer},\ }\href@noop {}
  {\bibfield  {journal} {\bibinfo  {journal} {Phys. Rev. Lett}\ }\textbf
  {\bibinfo {volume} {103}},\ \bibinfo {pages} {261801} (\bibinfo {year}
  {2009})}\BibitemShut {NoStop}%
\bibitem [{\citenamefont {Adelberger}\ \emph {et~al.}(2007)\citenamefont
  {Adelberger}, \citenamefont {Heckel}, \citenamefont {Hoedl}, \citenamefont
  {Hoyle}, \citenamefont {Kapner},\ and\ \citenamefont
  {Upadhye}}]{2006Particle}%
  \BibitemOpen
  \bibfield  {author} {\bibinfo {author} {\bibfnamefont {E.~G.}\ \bibnamefont
  {Adelberger}}, \bibinfo {author} {\bibfnamefont {B.~R.}\ \bibnamefont
  {Heckel}}, \bibinfo {author} {\bibfnamefont {S.}~\bibnamefont {Hoedl}},
  \bibinfo {author} {\bibfnamefont {C.~D.}\ \bibnamefont {Hoyle}}, \bibinfo
  {author} {\bibfnamefont {D.~J.}\ \bibnamefont {Kapner}},\ and\ \bibinfo
  {author} {\bibfnamefont {A.}~\bibnamefont {Upadhye}},\ }\bibfield  {title}
  {\bibinfo {title} {Particle physics implications of a recent test of the
  gravitational inverse square law},\ }\href@noop {} {\bibfield  {journal}
  {\bibinfo  {journal} {Phys.Rev.Lett}\ }\textbf {\bibinfo {volume} {98}},\
  \bibinfo {pages} {131104} (\bibinfo {year} {2007})}\BibitemShut {NoStop}%
\bibitem [{\citenamefont {Kapner}\ \emph {et~al.}(2007)\citenamefont {Kapner},
  \citenamefont {Cook}, \citenamefont {Adelberger}, \citenamefont {Gundlach},
  \citenamefont {Heckel}, \citenamefont {Hoyle},\ and\ \citenamefont
  {Swanson}}]{2007Tests1}%
  \BibitemOpen
  \bibfield  {author} {\bibinfo {author} {\bibfnamefont {D.~J.}\ \bibnamefont
  {Kapner}}, \bibinfo {author} {\bibfnamefont {T.~S.}\ \bibnamefont {Cook}},
  \bibinfo {author} {\bibfnamefont {E.~G.}\ \bibnamefont {Adelberger}},
  \bibinfo {author} {\bibfnamefont {J.~H.}\ \bibnamefont {Gundlach}}, \bibinfo
  {author} {\bibfnamefont {B.~R.}\ \bibnamefont {Heckel}}, \bibinfo {author}
  {\bibfnamefont {C.~D.}\ \bibnamefont {Hoyle}},\ and\ \bibinfo {author}
  {\bibfnamefont {H.~E.}\ \bibnamefont {Swanson}},\ }\bibfield  {title}
  {\bibinfo {title} {Tests of the gravitational inverse-square law below the
  dark-energy length scale},\ }\href@noop {} {\bibfield  {journal} {\bibinfo
  {journal} {Phys.Rev.Lett}\ }\textbf {\bibinfo {volume} {98}},\ \bibinfo
  {pages} {021101} (\bibinfo {year} {2007})}\BibitemShut {NoStop}%
\bibitem [{\citenamefont {Klimchitskaya}\ and\ \citenamefont
  {Mostepanenko}(2015)}]{2015Improved}%
  \BibitemOpen
  \bibfield  {author} {\bibinfo {author} {\bibfnamefont {G.~L.}\ \bibnamefont
  {Klimchitskaya}}\ and\ \bibinfo {author} {\bibfnamefont {V.~M.}\ \bibnamefont
  {Mostepanenko}},\ }\bibfield  {title} {\bibinfo {title} {Improved constraints
  on the coupling constants of axion-like particles to nucleons from recent
  casimir-less experiment},\ }\href@noop {} {\bibfield  {journal} {\bibinfo
  {journal} {Eur. Phys. J. C}\ }\textbf {\bibinfo {volume} {75}},\ \bibinfo
  {pages} {1} (\bibinfo {year} {2015})}\BibitemShut {NoStop}%
\bibitem [{\citenamefont {Chen}\ \emph {et~al.}(2016)\citenamefont {Chen},
  \citenamefont {Tham}, \citenamefont {Krause}, \citenamefont {Lopez},
  \citenamefont {Fischbach},\ and\ \citenamefont {DeCca}}]{2016Stronger}%
  \BibitemOpen
  \bibfield  {author} {\bibinfo {author} {\bibfnamefont {Y.~J.}\ \bibnamefont
  {Chen}}, \bibinfo {author} {\bibfnamefont {W.~K.}\ \bibnamefont {Tham}},
  \bibinfo {author} {\bibfnamefont {D.}~\bibnamefont {Krause}}, \bibinfo
  {author} {\bibfnamefont {D.}~\bibnamefont {Lopez}}, \bibinfo {author}
  {\bibfnamefont {E.}~\bibnamefont {Fischbach}},\ and\ \bibinfo {author}
  {\bibfnamefont {R.S.}\ \bibnamefont {Decca}},\ }\bibfield  {title}
  {\bibinfo {title} {Stronger limits on hypothetical yukawa interactions in the
  30–8000nm range},\ }\href@noop {} {\bibfield  {journal} {\bibinfo
  {journal} {Phys. Rev. Lett}\ }\textbf {\bibinfo {volume} {116}},\ \bibinfo
  {pages} {221102} (\bibinfo {year} {2016})}\BibitemShut {NoStop}%
\bibitem [{\citenamefont {Bezerra}\ \emph
  {et~al.}(2014{\natexlab{a}})\citenamefont {Bezerra}, \citenamefont
  {Klimchitskaya}, \citenamefont {Mostepanenko},\ and\ \citenamefont
  {Romero}}]{2014Constraining}%
  \BibitemOpen
  \bibfield  {author} {\bibinfo {author} {\bibfnamefont {V.~B.}\ \bibnamefont
  {Bezerra}}, \bibinfo {author} {\bibfnamefont {G.~L.}\ \bibnamefont
  {Klimchitskaya}}, \bibinfo {author} {\bibfnamefont {V.~M.}\ \bibnamefont
  {Mostepanenko}},\ and\ \bibinfo {author} {\bibfnamefont {C.}~\bibnamefont
  {Romero}},\ }\bibfield  {title} {\bibinfo {title} {Constraining axion-nucleon
  coupling constants from measurements of effective casimir pressure by means
  of micromachined oscillator},\ }\href@noop {} {\bibfield  {journal} {\bibinfo
   {journal} {Eur. Phys. J. C}\ }\textbf {\bibinfo {volume} {74}},\ \bibinfo
  {pages} {2859} (\bibinfo {year} {2014}{\natexlab{a}})}\BibitemShut {NoStop}%
\bibitem [{\citenamefont {Bezerra}\ \emph
  {et~al.}(2014{\natexlab{b}})\citenamefont {Bezerra}, \citenamefont
  {Klimchitskaya}, \citenamefont {Mostepanenko},\ and\ \citenamefont
  {Romero}}]{2014Constraints}%
  \BibitemOpen
  \bibfield  {author} {\bibinfo {author} {\bibfnamefont {V.~B.}\ \bibnamefont
  {Bezerra}}, \bibinfo {author} {\bibfnamefont {G.~L.}\ \bibnamefont
  {Klimchitskaya}}, \bibinfo {author} {\bibfnamefont {V.~M.}\ \bibnamefont
  {Mostepanenko}},\ and\ \bibinfo {author} {\bibfnamefont {C.}~\bibnamefont
  {Romero}},\ }\bibfield  {title} {\bibinfo {title} {Constraints on
  axion-nucleon coupling constants from measuring the casimir force between
  corrugated surfaces},\ }\href@noop {} {\bibfield  {journal} {\bibinfo
  {journal} {Phys.Rev.D}\ }\textbf {\bibinfo {volume} {90}},\ \bibinfo {pages}
  {055013} (\bibinfo {year} {2014}{\natexlab{b}})}\BibitemShut {NoStop}%
\bibitem [{\citenamefont {Klimchitskaya}\ and\ \citenamefont
  {Mostepanenko}(2017)}]{2017Constraints}%
  \BibitemOpen
  \bibfield  {author} {\bibinfo {author} {\bibfnamefont {G.~L.}\ \bibnamefont
  {Klimchitskaya}}\ and\ \bibinfo {author} {\bibfnamefont {V.~M.}\ \bibnamefont
  {Mostepanenko}},\ }\bibfield  {title} {\bibinfo {title} {Constraints on
  axion-like particles and non-newtonian gravity from measuring the difference
  of casimir forces},\ }\href@noop {} {\bibfield  {journal} {\bibinfo
  {journal} {Phys. Rev. D}\ }\textbf {\bibinfo {volume} {95}},\ \bibinfo
  {pages} {123013} (\bibinfo {year} {2017})}\BibitemShut {NoStop}%
\bibitem [{\citenamefont {Bezerra}\ \emph
  {et~al.}(2014{\natexlab{c}})\citenamefont {Bezerra}, \citenamefont
  {Klimchitskaya}, \citenamefont {Mostepanenko},\ and\ \citenamefont
  {Romero}}]{V2014Stronger}%
  \BibitemOpen
  \bibfield  {author} {\bibinfo {author} {\bibfnamefont {V.~B.}\ \bibnamefont
  {Bezerra}}, \bibinfo {author} {\bibfnamefont {G.}~\bibnamefont
  {Klimchitskaya}}, \bibinfo {author} {\bibfnamefont {V.}~\bibnamefont
  {Mostepanenko}},\ and\ \bibinfo {author} {\bibfnamefont {C.}~\bibnamefont
  {Romero}},\ }\bibfield  {title} {\bibinfo {title} {Stronger constraints on an
  axion from measuring the casimir interaction by means of a dynamic atomic
  force microscope},\ }\href@noop {} {\bibfield  {journal} {\bibinfo  {journal}
  {Phys Rev. D}\ }\textbf {\bibinfo {volume} {89}},\ \bibinfo {pages} {075002}
  (\bibinfo {year} {2014}{\natexlab{c}})}\BibitemShut {NoStop}%
\bibitem [{\citenamefont {Bezerra}\ \emph
  {et~al.}(2014{\natexlab{d}})\citenamefont {Bezerra}, \citenamefont
  {Klimchitskaya}, \citenamefont {Mostepanenko},\ and\ \citenamefont
  {Romero}}]{V2014Constraints}%
  \BibitemOpen
  \bibfield  {author} {\bibinfo {author} {\bibfnamefont {V.}~\bibnamefont
  {Bezerra}}, \bibinfo {author} {\bibfnamefont {G.}~\bibnamefont
  {Klimchitskaya}}, \bibinfo {author} {\bibfnamefont {V.}~\bibnamefont
  {Mostepanenko}},\ and\ \bibinfo {author} {\bibfnamefont {C.}~\bibnamefont
  {Romero}},\ }\bibfield  {title} {\bibinfo {title} {Constraints on the
  parameters of an axion from measurements of the thermal casimir-polder
  force},\ }\href@noop {} {\bibfield  {journal} {\bibinfo  {journal} {Phys.
  Rev. D}\ }\textbf {\bibinfo {volume} {89}},\ \bibinfo {pages} {035010}
  (\bibinfo {year} {2014}{\natexlab{d}})}\BibitemShut {NoStop}%
\bibitem [{\citenamefont {Decca}\ \emph {et~al.}(2007)\citenamefont {Decca},
  \citenamefont {López}, \citenamefont {Fischbach}, \citenamefont
  {Klimchitskaya}, \citenamefont {Krause},\ and\ \citenamefont
  {Mostepanenko}}]{2007Tests}%
  \BibitemOpen
  \bibfield  {author} {\bibinfo {author} {\bibfnamefont {R.~S.}\ \bibnamefont
  {Decca}}, \bibinfo {author} {\bibfnamefont {D.}~\bibnamefont {López}},
  \bibinfo {author} {\bibfnamefont {E.}~\bibnamefont {Fischbach}}, \bibinfo
  {author} {\bibfnamefont {G.~L.}\ \bibnamefont {Klimchitskaya}}, \bibinfo
  {author} {\bibfnamefont {D.~E.}\ \bibnamefont {Krause}},\ and\ \bibinfo
  {author} {\bibfnamefont {V.~M.}\ \bibnamefont {Mostepanenko}},\ }\bibfield
  {title} {\bibinfo {title} {Tests of new physics from precise measurements of
  the casimir pressure between two gold-coated plates},\ }\href@noop {}
  {\bibfield  {journal} {\bibinfo  {journal} {Phys. Rev. D}\ }\textbf {\bibinfo
  {volume} {75}},\ \bibinfo {pages} {077101} (\bibinfo {year}
  {2007})}\BibitemShut {NoStop}%
\bibitem [{\citenamefont {Decca}\ \emph {et~al.}(2010)\citenamefont {Decca},
  \citenamefont {Lopez}, \citenamefont {Fischbach}, \citenamefont
  {Klimchitskaya}, \citenamefont {Krause},\ and\ \citenamefont
  {Mostepanenko}}]{2010Novel}%
  \BibitemOpen
  \bibfield  {author} {\bibinfo {author} {\bibfnamefont {R.~S.}\ \bibnamefont
  {Decca}}, \bibinfo {author} {\bibfnamefont {D.}~\bibnamefont {Lopez}},
  \bibinfo {author} {\bibfnamefont {E.}~\bibnamefont {Fischbach}}, \bibinfo
  {author} {\bibfnamefont {G.~L.}\ \bibnamefont {Klimchitskaya}}, \bibinfo
  {author} {\bibfnamefont {D.~E.}\ \bibnamefont {Krause}},\ and\ \bibinfo
  {author} {\bibfnamefont {V.~M.}\ \bibnamefont {Mostepanenko}},\ }\bibfield
  {title} {\bibinfo {title} {Novel constraints on light elementary particles
  and extra-dimensional physics from the casimir effect},\ }\href@noop {}
  {\bibfield  {journal} {\bibinfo  {journal} {Eur. Phys. J. C}\ }\textbf
  {\bibinfo {volume} {51}},\ \bibinfo {pages} {963} (\bibinfo {year}
  {2010})}\BibitemShut {NoStop}%
\bibitem [{\citenamefont {Chiu}\ \emph {et~al.}(2009)\citenamefont {Chiu},
  \citenamefont {Klimchitskaya}, \citenamefont {Marachevsky}, \citenamefont
  {Mostepanenko},\ and\ \citenamefont {Mohideen}}]{2009Demonstration}%
  \BibitemOpen
  \bibfield  {author} {\bibinfo {author} {\bibfnamefont {H.~C.}\ \bibnamefont
  {Chiu}}, \bibinfo {author} {\bibfnamefont {G.~L.}\ \bibnamefont
  {Klimchitskaya}}, \bibinfo {author} {\bibfnamefont {V.~N.}\ \bibnamefont
  {Marachevsky}}, \bibinfo {author} {\bibfnamefont {V.~M.}\ \bibnamefont
  {Mostepanenko}},\ and\ \bibinfo {author} {\bibfnamefont {U.}~\bibnamefont
  {Mohideen}},\ }\bibfield  {title} {\bibinfo {title} {Demonstration of the
  asymmetric lateral casimir force between corrugated surfaces in the
  nonadditive regime},\ }\href@noop {} {\bibfield  {journal} {\bibinfo
  {journal} {Phys. Rev. B}\ }\textbf {\bibinfo {volume} {80}},\ \bibinfo
  {pages} {121402(R)} (\bibinfo {year} {2009})}\BibitemShut {NoStop}%
\bibitem [{\citenamefont {Chiu}\ \emph {et~al.}(2010)\citenamefont {Chiu},
  \citenamefont {Klimchitskaya}, \citenamefont {Marachevsky}, \citenamefont
  {Mostepanenko},\ and\ \citenamefont {Mohideen}}]{2010Lateral}%
  \BibitemOpen
  \bibfield  {author} {\bibinfo {author} {\bibfnamefont {H.~C.}\ \bibnamefont
  {Chiu}}, \bibinfo {author} {\bibfnamefont {G.~L.}\ \bibnamefont
  {Klimchitskaya}}, \bibinfo {author} {\bibfnamefont {V.~N.}\ \bibnamefont
  {Marachevsky}}, \bibinfo {author} {\bibfnamefont {V.~M.}\ \bibnamefont
  {Mostepanenko}},\ and\ \bibinfo {author} {\bibfnamefont {U.}~\bibnamefont
  {Mohideen}},\ }\bibfield  {title} {\bibinfo {title} {Lateral casimir force
  between sinusoidally corrugated surfaces: Asymmetric profiles, deviations
  from the proximity force approximation and comparison with exact theory},\
  }\href@noop {} {\bibfield  {journal} {\bibinfo  {journal} {Phys. Rev. B}\
  }\textbf {\bibinfo {volume} {81}},\ \bibinfo {pages} {760} (\bibinfo {year}
  {2010})}\BibitemShut {NoStop}%
\bibitem [{\citenamefont {Bimonte}\ \emph {et~al.}(2016)\citenamefont
  {Bimonte}, \citenamefont {Lopez},\ and\ \citenamefont
  {Decca}}]{2016Isoelectronic}%
  \BibitemOpen
  \bibfield  {author} {\bibinfo {author} {\bibfnamefont {G.}~\bibnamefont
  {Bimonte}}, \bibinfo {author} {\bibfnamefont {D.}~\bibnamefont {Lopez}},\
  and\ \bibinfo {author} {\bibfnamefont {R.~S.}\ \bibnamefont {Decca}},\
  }\bibfield  {title} {\bibinfo {title} {Isoelectronic determination of the
  thermal casimir force},\ }\href@noop {} {\bibfield  {journal} {\bibinfo
  {journal} {Phys. Rev. B}\ }\textbf {\bibinfo {volume} {93}},\ \bibinfo
  {pages} {184434} (\bibinfo {year} {2016})}\BibitemShut {NoStop}%
\bibitem [{\citenamefont {Chang}\ \emph {et~al.}(2012)\citenamefont {Chang},
  \citenamefont {Banishev}, \citenamefont {Castillo-Garza}, \citenamefont
  {Klimchitskaya}, \citenamefont {Mostepanenko},\ and\ \citenamefont
  {Mohideen}}]{2012Gradient}%
  \BibitemOpen
  \bibfield  {author} {\bibinfo {author} {\bibfnamefont {C.}~\bibnamefont
  {Chang}}, \bibinfo {author} {\bibfnamefont {A.~A.}\ \bibnamefont {Banishev}},
  \bibinfo {author} {\bibfnamefont {R.}~\bibnamefont {Castillo-Garza}},
  \bibinfo {author} {\bibfnamefont {G.~L.}\ \bibnamefont {Klimchitskaya}},
  \bibinfo {author} {\bibfnamefont {V.}~\bibnamefont {Mostepanenko}},\ and\
  \bibinfo {author} {\bibfnamefont {U.}~\bibnamefont {Mohideen}},\ }\bibfield
  {title} {\bibinfo {title} {Gradient of the casimir force between au surfaces
  of a sphere and a plate measured using an atomic force microscope in a
  frequency-shift technique},\ }\href@noop {} {\bibfield  {journal} {\bibinfo
  {journal} {Phys. Rev. B}\ }\textbf {\bibinfo {volume} {85}},\ \bibinfo
  {pages} {543} (\bibinfo {year} {2012})}\BibitemShut {NoStop}%
\bibitem [{\citenamefont {Obrecht}\ \emph {et~al.}(2007)\citenamefont
  {Obrecht}, \citenamefont {Wild}, \citenamefont {Antezza}, \citenamefont
  {Pitaevskii}, \citenamefont {Stringari},\ and\ \citenamefont
  {Cornell}}]{2007Measurement}%
  \BibitemOpen
  \bibfield  {author} {\bibinfo {author} {\bibfnamefont {J.~M.}\ \bibnamefont
  {Obrecht}}, \bibinfo {author} {\bibfnamefont {R.~J.}\ \bibnamefont {Wild}},
  \bibinfo {author} {\bibfnamefont {M.}~\bibnamefont {Antezza}}, \bibinfo
  {author} {\bibfnamefont {L.~P.}\ \bibnamefont {Pitaevskii}}, \bibinfo
  {author} {\bibfnamefont {S.}~\bibnamefont {Stringari}},\ and\ \bibinfo
  {author} {\bibfnamefont {E.~A.}\ \bibnamefont {Cornell}},\ }\bibfield
  {title} {\bibinfo {title} {Measurement of the temperature dependence of the
  casimir-polder force},\ }\href@noop {} {\bibfield  {journal} {\bibinfo
  {journal} {Phys. Rev. Lett}\ }\textbf {\bibinfo {volume} {98}},\ \bibinfo
  {pages} {063201} (\bibinfo {year} {2007})}\BibitemShut {NoStop}%
\bibitem [{\citenamefont {Sedighi}\ \emph {et~al.}(2016)\citenamefont
  {Sedighi}, \citenamefont {Svetovoy},\ and\ \citenamefont
  {Palasantzas}}]{PhysRevB.93.085434}%
  \BibitemOpen
  \bibfield  {author} {\bibinfo {author} {\bibfnamefont {M.}~\bibnamefont
  {Sedighi}}, \bibinfo {author} {\bibfnamefont {V.~B.}\ \bibnamefont
  {Svetovoy}},\ and\ \bibinfo {author} {\bibfnamefont {G.}~\bibnamefont
  {Palasantzas}},\ }\bibfield  {title} {\bibinfo {title} {Casimir force
  measurements from silicon carbide surfaces},\ }\href
  {https://doi.org/10.1103/PhysRevB.93.085434} {\bibfield  {journal} {\bibinfo
  {journal} {Phys. Rev. B}\ }\textbf {\bibinfo {volume} {93}},\ \bibinfo
  {pages} {085434} (\bibinfo {year} {2016})}\BibitemShut {NoStop}%
\bibitem [{\citenamefont {Ramsey}(1979)}]{1979The}%
  \BibitemOpen
  \bibfield  {author} {\bibinfo {author} {\bibfnamefont {N.~F.}\ \bibnamefont
  {Ramsey}},\ }\bibfield  {title} {\bibinfo {title} {The tensor force between
  two protons at long range},\ }\href@noop {} {\bibfield  {journal} {\bibinfo
  {journal} {Physica A Statistical Mechanics Its Applications}\ }\textbf
  {\bibinfo {volume} {96}},\ \bibinfo {pages} {285} (\bibinfo {year}
  {1979})}\BibitemShut {NoStop}%
\bibitem [{\citenamefont {Ledbetter}\ \emph {et~al.}(2013)\citenamefont
  {Ledbetter}, \citenamefont {Romalis},\ and\ \citenamefont
  {Kimball}}]{2013Constraints}%
  \BibitemOpen
  \bibfield  {author} {\bibinfo {author} {\bibfnamefont {M.~P.}\ \bibnamefont
  {Ledbetter}}, \bibinfo {author} {\bibfnamefont {M.~V.}\ \bibnamefont
  {Romalis}},\ and\ \bibinfo {author} {\bibfnamefont {D.}~\bibnamefont
  {Kimball}},\ }\bibfield  {title} {\bibinfo {title} {Constraints on
  short-range spin-dependent interactions from scalar spin-spin coupling in
  deuterated molecular hydrogen},\ }\href@noop {} {\bibfield  {journal}
  {\bibinfo  {journal} {Phys. Rev. Lett}\ }\textbf {\bibinfo {volume} {110}},\
  \bibinfo {pages} {040402} (\bibinfo {year} {2013})}\BibitemShut {NoStop}%
\bibitem [{\citenamefont {Genes}\ \emph {et~al.}(2008)\citenamefont {Genes},
  \citenamefont {Vitali}, \citenamefont {Tombesi}, \citenamefont {Gigan},\ and\
  \citenamefont {Aspelmeyer}}]{2008Ground}%
  \BibitemOpen
  \bibfield  {author} {\bibinfo {author} {\bibfnamefont {C.}~\bibnamefont
  {Genes}}, \bibinfo {author} {\bibfnamefont {D.}~\bibnamefont {Vitali}},
  \bibinfo {author} {\bibfnamefont {P.}~\bibnamefont {Tombesi}}, \bibinfo
  {author} {\bibfnamefont {S.}~\bibnamefont {Gigan}},\ and\ \bibinfo {author}
  {\bibfnamefont {M.}~\bibnamefont {Aspelmeyer}},\ }\bibfield  {title}
  {\bibinfo {title} {Ground-state cooling of a micromechanical oscillator:
  Comparing cold damping and cavity-assisted cooling schemes},\ }\href@noop {}
  {\bibfield  {journal} {\bibinfo  {journal} {Seminars in Oncology Nursing}\
  }\textbf {\bibinfo {volume} {7}},\ \bibinfo {pages} {26} (\bibinfo {year}
  {2008})}\BibitemShut {NoStop}%
\bibitem [{\citenamefont {Yin}\ \emph {et~al.}(2013)\citenamefont {Yin},
  \citenamefont {Geraci},\ and\ \citenamefont {Li}}]{2013Optomechanics}%
  \BibitemOpen
  \bibfield  {author} {\bibinfo {author} {\bibfnamefont {Z.~Q.}\ \bibnamefont
  {Yin}}, \bibinfo {author} {\bibfnamefont {A.~A.}\ \bibnamefont {Geraci}},\
  and\ \bibinfo {author} {\bibfnamefont {T.}~\bibnamefont {Li}},\ }\bibfield
  {title} {\bibinfo {title} {Optomechanics of levitated dielectric particles},\
  }\href@noop {} {\bibfield  {journal} {\bibinfo  {journal} {International
  Journal of Modern Physics B}\ }\textbf {\bibinfo {volume} {27}},\ \bibinfo
  {pages} {1330018} (\bibinfo {year} {2013})}\BibitemShut {NoStop}%
\bibitem [{\citenamefont {Millen}\ \emph {et~al.}(2020)\citenamefont {Millen},
  \citenamefont {Monteiro}, \citenamefont {Pettit},\ and\ \citenamefont
  {Vamivakas}}]{2020Optomechanics}%
  \BibitemOpen
  \bibfield  {author} {\bibinfo {author} {\bibfnamefont {J.}~\bibnamefont
  {Millen}}, \bibinfo {author} {\bibfnamefont {T.~S.}\ \bibnamefont
  {Monteiro}}, \bibinfo {author} {\bibfnamefont {R.}~\bibnamefont {Pettit}},\
  and\ \bibinfo {author} {\bibfnamefont {A.~N.}\ \bibnamefont {Vamivakas}},\
  }\bibfield  {title} {\bibinfo {title} {Optomechanics with levitated
  particles},\ }\href@noop {} {\bibfield  {journal} {\bibinfo  {journal} {Rep.
  Prog. Phys}\ }\textbf {\bibinfo {volume} {83}} (\bibinfo {year}
  {2020})}\BibitemShut {NoStop}%
\bibitem [{\citenamefont {Weis}\ \emph {et~al.}()\citenamefont {Weis},
  \citenamefont {Riviere}, \citenamefont {Deleglise}, \citenamefont {Gavartin},
  \citenamefont {Arcizet}, \citenamefont {Schliesser},\ and\ \citenamefont
  {Kippenberg}}]{2010Optomechanically}%
  \BibitemOpen
  \bibfield  {author} {\bibinfo {author} {\bibfnamefont {S.}~\bibnamefont
  {Weis}}, \bibinfo {author} {\bibfnamefont {R.}~\bibnamefont {Riviere}},
  \bibinfo {author} {\bibfnamefont {S.}~\bibnamefont {Deleglise}}, \bibinfo
  {author} {\bibfnamefont {E.}~\bibnamefont {Gavartin}}, \bibinfo {author}
  {\bibfnamefont {O.}~\bibnamefont {Arcizet}}, \bibinfo {author} {\bibfnamefont
  {A.}~\bibnamefont {Schliesser}},\ and\ \bibinfo {author} {\bibfnamefont
  {T.}~\bibnamefont {Kippenberg}},\ }\bibfield  {title} {\bibinfo {title}
  {Optomechanically induced transparency.},\ }\href@noop {} {\bibfield
  {journal} {\bibinfo  {journal} {Science}\ }\textbf {\bibinfo {volume}
  {330}},\ \bibinfo {pages} {1520}}\BibitemShut {NoStop}%
\bibitem [{\citenamefont {Liu}\ \emph {et~al.}(2019)\citenamefont {Liu},
  \citenamefont {Xu}, \citenamefont {Klimchitskaya}, \citenamefont
  {Mostepanenko},\ and\ \citenamefont {Mohideen}}]{2019Precision}%
  \BibitemOpen
  \bibfield  {author} {\bibinfo {author} {\bibfnamefont {M.}~\bibnamefont
  {Liu}}, \bibinfo {author} {\bibfnamefont {J.}~\bibnamefont {Xu}}, \bibinfo
  {author} {\bibfnamefont {G.~L.}\ \bibnamefont {Klimchitskaya}}, \bibinfo
  {author} {\bibfnamefont {V.~M.}\ \bibnamefont {Mostepanenko}},\ and\ \bibinfo
  {author} {\bibfnamefont {U.}~\bibnamefont {Mohideen}},\ }\bibfield  {title}
  {\bibinfo {title} {Precision measurements of the gradient of the casimir
  force between ultra clean metallic surfaces at larger separations},\
  }\href@noop {} {\bibfield  {journal} {\bibinfo  {journal} {Phys. Rev. A}\
  }\textbf {\bibinfo {volume} {100}},\ \bibinfo {pages} {052511} (\bibinfo
  {year} {2019})}\BibitemShut {NoStop}%
\bibitem [{\citenamefont {Giessibl∗}(2003)}]{Franz2003Advances}%
  \BibitemOpen
  \bibfield  {author} {\bibinfo {author} {\bibfnamefont {F.~J.}\ \bibnamefont
  {Giessibl}},\ }\bibfield  {title} {\bibinfo {title} {Advances in atomic
  force microscopy},\ }\href@noop {} {\bibfield  {journal} {\bibinfo  {journal}
  {Rev. Mod. Phys}\ }\textbf {\bibinfo {volume} {75}} (\bibinfo {year}
  {2003})}\BibitemShut {NoStop}%
\bibitem [{\citenamefont {Decca}\ \emph {et~al.}(2005)\citenamefont {Decca},
  \citenamefont {Lopez}, \citenamefont {Chan}, \citenamefont {Fischbach},
  \citenamefont {Krause},\ and\ \citenamefont {Jamell}}]{2005Constraining}%
  \BibitemOpen
  \bibfield  {author} {\bibinfo {author} {\bibfnamefont {R.~S.}\ \bibnamefont
  {Decca}}, \bibinfo {author} {\bibfnamefont {D.}~\bibnamefont {Lopez}},
  \bibinfo {author} {\bibfnamefont {H.~B.}\ \bibnamefont {Chan}}, \bibinfo
  {author} {\bibfnamefont {E.}~\bibnamefont {Fischbach}}, \bibinfo {author}
  {\bibfnamefont {D.~E.}\ \bibnamefont {Krause}},\ and\ \bibinfo {author}
  {\bibfnamefont {C.~R.}\ \bibnamefont {Jamell}},\ }\bibfield  {title}
  {\bibinfo {title} {Constraining new forces in the casimir regime using the
  isoelectronic technique},\ }\href@noop {} {\bibfield  {journal} {\bibinfo
  {journal} {Phys. Rev. Lett}\ }\textbf {\bibinfo {volume} {94}},\ \bibinfo
  {pages} {240401} (\bibinfo {year} {2005})}\BibitemShut {NoStop}%
\bibitem [{\citenamefont {Adelberger}\ \emph {et~al.}(2003)\citenamefont
  {Adelberger}, \citenamefont {Fischbach}, \citenamefont {Krause},\ and\
  \citenamefont {Newman}}]{2003Constraining}%
  \BibitemOpen
  \bibfield  {author} {\bibinfo {author} {\bibfnamefont {E.~G.}\ \bibnamefont
  {Adelberger}}, \bibinfo {author} {\bibfnamefont {E.}~\bibnamefont
  {Fischbach}}, \bibinfo {author} {\bibfnamefont {D.~E.}\ \bibnamefont
  {Krause}},\ and\ \bibinfo {author} {\bibfnamefont {R.~D.}\ \bibnamefont
  {Newman}},\ }\bibfield  {title} {\bibinfo {title} {Constraining the couplings
  of massive pseudoscalars using gravity and optical experiments},\ }\href@noop
  {} {\bibfield  {journal} {\bibinfo  {journal} {Phys. Rev. D}\ }\textbf
  {\bibinfo {volume} {68}},\ \bibinfo {pages} {062002} (\bibinfo {year}
  {2003})}\BibitemShut {NoStop}%
\bibitem [{\citenamefont {Chang}\ \emph {et~al.}(2009)\citenamefont {Chang},
  \citenamefont {Regal}, \citenamefont {Papp}, \citenamefont {Wilson},\ and\
  \citenamefont {Ye}}]{2009Cavity}%
  \BibitemOpen
  \bibfield  {author} {\bibinfo {author} {\bibfnamefont {D.~E.}\ \bibnamefont
  {Chang}}, \bibinfo {author} {\bibfnamefont {C.~A.}\ \bibnamefont {Regal}},
  \bibinfo {author} {\bibfnamefont {S.~B.}\ \bibnamefont {Papp}}, \bibinfo
  {author} {\bibfnamefont {D.~J.}\ \bibnamefont {Wilson}},\ and\ \bibinfo
  {author} {\bibfnamefont {J.}~\bibnamefont {Ye}},\ }\bibfield  {title}
  {\bibinfo {title} {Cavity opto-mechanics using an optically levitated
  nanosphere},\ }\href@noop {} {\bibfield  {journal} {\bibinfo  {journal}
  {Proceedings of the National Academy of Sciences of the United States of
  America}\ }\textbf {\bibinfo {volume} {107}},\ \bibinfo {pages} {1005}
  (\bibinfo {year} {2009})}\BibitemShut {NoStop}%
\bibitem [{\citenamefont {Grblacher}\ \emph {et~al.}(2009)\citenamefont
  {Grblacher}, \citenamefont {Hammerer}, \citenamefont {Vanner},\ and\
  \citenamefont {Aspelmeyer}}]{2009Observation}%
  \BibitemOpen
  \bibfield  {author} {\bibinfo {author} {\bibfnamefont {S.}~\bibnamefont
  {Grblacher}}, \bibinfo {author} {\bibfnamefont {K.}~\bibnamefont {Hammerer}},
  \bibinfo {author} {\bibfnamefont {M.~R.}\ \bibnamefont {Vanner}},\ and\
  \bibinfo {author} {\bibfnamefont {M.}~\bibnamefont {Aspelmeyer}},\ }\bibfield
   {title} {\bibinfo {title} {Observation of strong coupling between a
  micromechanical resonator and an optical cavity field},\ }\href@noop {}
  {\bibfield  {journal} {\bibinfo  {journal} {Nature}\ }\textbf {\bibinfo
  {volume} {460}},\ \bibinfo {pages} {724} (\bibinfo {year}
  {2009})}\BibitemShut {NoStop}%
\bibitem [{\citenamefont {Cleland}\ and\ \citenamefont
  {Roukes}(2002)}]{2002Noise}%
  \BibitemOpen
  \bibfield  {author} {\bibinfo {author} {\bibfnamefont {A.~N.}\ \bibnamefont
  {Cleland}}\ and\ \bibinfo {author} {\bibfnamefont {M.~L.}\ \bibnamefont
  {Roukes}},\ }\bibfield  {title} {\bibinfo {title} {Noise processes in
  nanomechanical resonators},\ }\href@noop {} {\bibfield  {journal} {\bibinfo
  {journal} {Journal of Applied Physics}\ }\textbf {\bibinfo {volume} {92}},\
  \bibinfo {pages} {2758} (\bibinfo {year} {2002})}\BibitemShut {NoStop}%
\bibitem [{\citenamefont {Ekinci}\ \emph {et~al.}(2004)\citenamefont {Ekinci},
  \citenamefont {Yang},\ and\ \citenamefont {Roukes}}]{Ekinci2004Ultimate}%
  \BibitemOpen
  \bibfield  {author} {\bibinfo {author} {\bibfnamefont {K.~L.}\ \bibnamefont
  {Ekinci}}, \bibinfo {author} {\bibfnamefont {Y.}~\bibnamefont {Yang}},\ and\
  \bibinfo {author} {\bibfnamefont {M.~L.}\ \bibnamefont {Roukes}},\ }\bibfield
   {title} {\bibinfo {title} {Ultimate limits to inertial mass sensing based
  upon nanoelectromechanical systems},\ }\href@noop {} {\bibfield  {journal}
  {\bibinfo  {journal} {Journal of Applied Physics}\ }\textbf {\bibinfo
  {volume} {95}},\ \bibinfo {pages} {2682} (\bibinfo {year}
  {2004})}\BibitemShut {NoStop}%
\bibitem [{\citenamefont {Goryachev}\ \emph {et~al.}(2018)\citenamefont
  {Goryachev}, \citenamefont {Mcallister},\ and\ \citenamefont
  {Tobar}}]{2018Probing}%
  \BibitemOpen
  \bibfield  {author} {\bibinfo {author} {\bibfnamefont {M.}~\bibnamefont
  {Goryachev}}, \bibinfo {author} {\bibfnamefont {B.}~\bibnamefont
  {Mcallister}},\ and\ \bibinfo {author} {\bibfnamefont {M.~E.}\ \bibnamefont
  {Tobar}},\ }\bibfield  {title} {\bibinfo {title} {Probing dark universe with
  exceptional points},\ }\href@noop {} {\bibfield  {journal} {\bibinfo
  {journal} {Physics of the Dark Universe}\ }\textbf {\bibinfo {volume} {23}}
  (\bibinfo {year} {2018})}\BibitemShut {NoStop}%
\bibitem [{\citenamefont {Vollmer}\ and\ \citenamefont
  {Yang}(2012)}]{2012Review}%
  \BibitemOpen
  \bibfield  {author} {\bibinfo {author} {\bibfnamefont {F.}~\bibnamefont
  {Vollmer}}\ and\ \bibinfo {author} {\bibfnamefont {L.}~\bibnamefont {Yang}},\
  }\bibfield  {title} {\bibinfo {title} {Review label-free detection with
  high-q microcavities: a review of biosensing mechanisms for integrated
  devices},\ }\href@noop {} {\bibfield  {journal} {\bibinfo  {journal}
  {Nanophotonics}\ }\textbf {\bibinfo {volume} {1}},\ \bibinfo {pages} {267}
  (\bibinfo {year} {2012})}\BibitemShut {NoStop}%
\bibitem [{\citenamefont {Fischbach}\ \emph {et~al.}(2003)\citenamefont
  {Fischbach}, \citenamefont {Krause}, \citenamefont {Decca},\ and\
  \citenamefont {López}}]{2003Testing}%
  \BibitemOpen
  \bibfield  {author} {\bibinfo {author} {\bibfnamefont {E.}~\bibnamefont
  {Fischbach}}, \bibinfo {author} {\bibfnamefont {D.~E.}\ \bibnamefont
  {Krause}}, \bibinfo {author} {\bibfnamefont {R.~S.}\ \bibnamefont {Decca}},\
  and\ \bibinfo {author} {\bibfnamefont {D.}~\bibnamefont {López}},\
  }\bibfield  {title} {\bibinfo {title} {Testing newtonian gravity at the
  nanometer distance scale using the iso-electronic effect},\ }\href@noop {}
  {\bibfield  {journal} {\bibinfo  {journal} {Physics Letters A}\ }\textbf
  {\bibinfo {volume} {318}},\ \bibinfo {pages} {165} (\bibinfo {year}
  {2003})}\BibitemShut {NoStop}%
\bibitem [{\citenamefont {Bezerra}\ \emph {et~al.}(2010)\citenamefont
  {Bezerra}, \citenamefont {Klimchitskaya}, \citenamefont {Mostepanenko},\ and\
  \citenamefont {Romero}}]{2010Advance}%
  \BibitemOpen
  \bibfield  {author} {\bibinfo {author} {\bibfnamefont {V.~B.}\ \bibnamefont
  {Bezerra}}, \bibinfo {author} {\bibfnamefont {G.~L.}\ \bibnamefont
  {Klimchitskaya}}, \bibinfo {author} {\bibfnamefont {V.~M.}\ \bibnamefont
  {Mostepanenko}},\ and\ \bibinfo {author} {\bibfnamefont {C.}~\bibnamefont
  {Romero}},\ }\bibfield  {title} {\bibinfo {title} {Advance and prospects in
  constraining the yukawa-type corrections to newtonian gravity from the
  casimir effect},\ }\href@noop {} {\bibfield  {journal} {\bibinfo  {journal}
  {Phys. Rev. D}\ }\textbf {\bibinfo {volume} {81}},\ \bibinfo {pages} {211}
  (\bibinfo {year} {2010})}\BibitemShut {NoStop}%
\bibitem [{\citenamefont {Geraci}\ \emph {et~al.}(2010)\citenamefont {Geraci},
  \citenamefont {Papp},\ and\ \citenamefont {Kitching}}]{2010Short}%
  \BibitemOpen
  \bibfield  {author} {\bibinfo {author} {\bibfnamefont {A.~A.}\ \bibnamefont
  {Geraci}}, \bibinfo {author} {\bibfnamefont {S.~B.}\ \bibnamefont {Papp}},\
  and\ \bibinfo {author} {\bibfnamefont {J.}~\bibnamefont {Kitching}},\
  }\bibfield  {title} {\bibinfo {title} {Short-range force detection using
  optically-cooled levitated microspheres},\ }\href@noop {} {\bibfield
  {journal} {\bibinfo  {journal} {Phys. Rev. Lett}\ }\textbf {\bibinfo {volume}
  {105}},\ \bibinfo {pages} {101101} (\bibinfo {year} {2010})}\BibitemShut
  {NoStop}%
\end{thebibliography}%

\end{document}